\documentclass[aps, a4paper, twocolumn, nofootinbib, showpacs, amsfonts, superscriptaddress]{revtex4}
\usepackage[utf8]{inputenc}
\usepackage{ dsfont }
\usepackage{amssymb}

\usepackage{graphicx}
\usepackage{amsmath}
\usepackage{bm}
\usepackage{xcolor}
\usepackage{tikz}
\usepackage{mathrsfs}
\usepackage{soul}
\usetikzlibrary{arrows}

\newcommand{\bea}{\begin{eqnarray}}
\newcommand{\eea}{\end{eqnarray}}
\newcommand{\be}{\begin{equation}}
\newcommand{\ee}{\end{equation}}
\newcommand{\nn}{\nonumber}

\begin{document}

\title{An emergent cosmological model from running Newton constant}

\author{Aknur Zholdasbek}
\email{aknur.zholdasbek@nu.edu.kz}
\affiliation{Department of Physics, Nazarbayev University, Kabanbay Batyr 53, 010000 Astana, Kazakhstan}

\author{Hrishikesh Chakrabarty}
\email{hrishikesh.chakrabarty@nu.edu.kz }
\affiliation{Department of Physics, Nazarbayev University, Kabanbay Batyr 53, 010000 Astana, Kazakhstan}

\author{Daniele Malafarina}
\email{daniele.malafarina@nu.edu.kz}
\affiliation{Department of Physics, Nazarbayev University, Kabanbay Batyr 53, 010000 Astana, Kazakhstan}

\author{Alfio Bonanno}
\email{alfio.bonanno@inaf.it}
\affiliation{INAF, Osservatorio Astrofisico di Catania, via S.Sofia 78, I-95123 Catania, Italy}

\begin{abstract}
We propose an emergent cosmological model rooted in the Asymptotically Safe antiscreening behavior of the Newton constant at Planckian energies. Distinguishing itself from prior approaches, our model encapsulates the variable nature of $G$ through a multiplicative coupling within the matter Lagrangian, characterized by a conserved energy-momentum tensor. The universe emerges from a quasi-de Sitter phase, transitioning to standard cosmological evolution post-Planck Era. Our analysis demonstrates the feasibility of constraining the transition scale to nearly classical cosmology using Cosmic Microwave Background (CMB) data and the potential to empirically probe the antiscreening trait of Newton's constant, as predicted by Asymptotic Safety. 
\end{abstract}

\maketitle

\section{Introduction}
For decades, the standard model of cosmology has been very successful in describing the observable universe in accordance with experimental observations. 
We are currently living in an era of precision cosmology in which observations of the cosmic microwave background (CMB) \cite{WMAP:2012nax, Planck:2018jri,Planck:2018nkj,Planck:2018vyg,Planck:2019nip}, large scale structure (LSS) \cite{SDSS:2003eyi,SDSS:2004kqt}, type-IA supernovae (SNeIa) \cite{SupernovaCosmologyProject:1998vns,SupernovaSearchTeam:1998fmf}, baryon acoustic oscillation (BAO) \cite{SDSS:2005xqv}, weak lensing \cite{Jain:2003tba}, gravitational waves (GW) \cite{LIGOScientific:2017adf} and more have allowed us to refine our picture of how the universe evolved by formulating, analyzing and potentially discarding theoretical models. 

However, it is becoming increasingly clear that the picture is not complete and there are still important open problems and issues that need to be addressed. Among these the idea of inflation, although extremely successful, has recently been put under scrutiny. According to inflation, the universe is believed to have gone through a period of accelerated expansion at very early times \cite{Guth:1980zm,Linde:1981mu,Starobinsky:1980te}. This idea was proposed to address two problems, known as the horizon and flatness problems, that plagued the early theory. 
Besides solving two major problems, the success of the inflationary paradigm relies on modern observations that show how the initial conditions for the hot big bang can be produced during this period of inflation.
The main breakthrough came when early CMB observations \cite{WMAP:2012nax} detected a nearly scale-invariant power spectrum which can be perfectly explained by inflation. Later observations from Planck made constraints on the model very strong, thus allowing to discard inflationary models that do not fit the observations. The latest Planck data favors single-field scalar models of inflation with concave potential coupled to General Relativity (GR). The data also prefers no significant departure from slow-roll behavior \cite{Planck:2018jri}.      
Still, the nature of the inflationary scalar field is not known at present. Therefore other models for inflation and the behavior of the universe in the early stages have been proposed over the years. Some of these models are able to address the same issue and produce a scale-invariant CMB power spectrum. For example, bouncing scenarios and ekpyrotic models have been suggested as possible alternatives to inflation \cite{Cai:2009zp, Battefeld:2014uga,Lyth:2001nq,Brandenberger:2016vhg}.
This shows that there is room for other viable models of the universe to be proposed and there are stringent quantitative tests that can be made to constrain such proposals.

In the present article, we propose a minimal modification to GR at UV scales inspired by the Asymptotic Safety (AS) paradigm \cite{Bonanno2023} and implemented through a multiplicative coupling of matter to gravity as proposed in \cite{Markov:1985py}. This proposal provides a relation between a variable Newton coupling $G$ and variable $\Lambda$ and does not involve any additional unknown fields. It is worth noting that the action proposed in \cite{Markov:1985py} may be viewed as part of a broader context of alternative theories with modifications of the matter Lagrangian $\mathcal{L}_m$ such as $f(\mathcal{L}_m)$ or $f(R,T)$ (where $T$ is the trace of the energy momentum tensor). 
Interested readers may refer to \cite{Harko:2010mv,Lobo:2012af,Harko:2011kv,Moraes:2017zgm,Bertolami:2008ab,Harko:2018gxr,Xu:2019sbp} and references therein.

We show that the quasi-de Sitter phase in the early universe can be obtained as a result of the Asymptotically Safe nature of gravity at high energies. Furthermore, we show that this new mechanism achieves the same results as scalar field inflation and we provide arguments in support of the fact that it can also produce a nearly scale-invariant primordial power spectrum.

The current proposal diverges from prior investigations into Renormalization Group (RG) improved cosmologies \cite{Bonanno:2001xi,Bonanno:2001hi,Cai:2011kd} by adopting a Lagrangian approach, thus preserving diffeomorphism invariance. While it shares conceptual similarities with the framework introduced in \cite{Bonanno:2012jy}, and subsequently elaborated upon in various studies (cf. \cite{Bonanno:2017pkg} for an overview), it avoids imposing a cutoff identification based on curvature invariants \cite{Bonanno:2018gck}. Recent findings, notably those derived from the fluctuation approach \cite{Pawlowski:2020qer,Bonanno:2021squ}, illustrate that the behavior of the Newton constant during high-energy scattering processes is predominantly governed by the external graviton momenta which should scale in accordance with the energy density of the system. A promising initial result stemming from this novel perspective is evidenced in the regular black hole collapse model discussed in \cite{Bonanno:2023rzk}.

The article is organized in the following way: In section \ref{AS} we briefly discuss AS gravity with the non trivial matter-gravity coupling and apply the formalism to a Friedmann universe in section \ref{cosmology}. Implications of the model for the early universe are discussed in section \ref{earlyU}. Finally results and future perspectives are summarized in section \ref{summary}. Throughout the article, we make use of units with $c=1$ (unless explicitly stated) and metric signature $(-, +, +, +)$.

\section{Field equations and energy momentum}\label{AS}

\begin{figure*}[]
    \begin{center}
        \includegraphics[width=8.5cm]{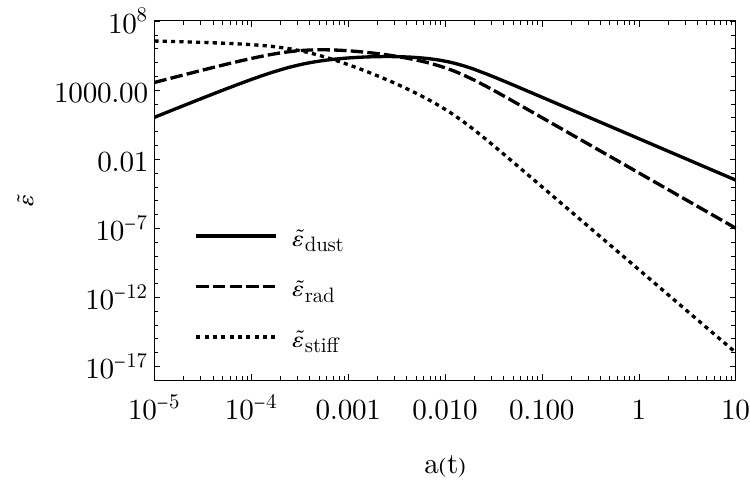}
        \includegraphics[width=8cm]{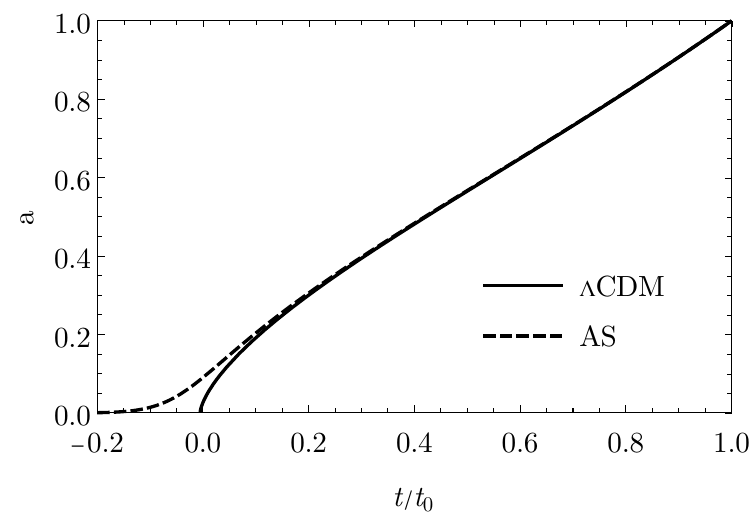}
    \end{center}
    \caption{Left: Behavior of the effective energy density $\Tilde{\varepsilon}$ for dust ($w=0$), radiation ($w=1/3$) and stiff fluid ($w=1$) in AS with MM coupling as a function of the scale factor $a$ of the universe, with the current epoch given by $a(t_0)=1$ and energy density $\varepsilon(t_0)=\varepsilon_{\rm 0}$. Here we take $\varepsilon_c = 10^{-5}$, $\varepsilon_{\rm 0,dust} = 0.3$, $\varepsilon_{\rm 0,rad} = 10^{-3}$ and $\varepsilon_{\rm 0,stiff} = 10^{-10}$ for illustrative purposes. Right: Behavior of the scale factor $ a $ with respect to $t/t_0$, where $ t_0 = 13.8 \ {\rm Gyr} $ is the age of the universe today, in the $\Lambda$-cdm model (solid line) as compared to the AS with MM coupling model (dashed line). For this plot we use $ \varepsilon_{k0}/\varepsilon_c = 10^{-2} $ for illustrative purposes and set the fiducial cosmological parameters from \cite{Planck:2018vyg}. \label{fig-loglog}}
\end{figure*}

We aim to build a cosmological model based on three main assumptions regarding the fundamental behavior of gravity at high energies:
\begin{itemize}
    \item[(i)] There exists a non-minimal matter-gravity coupling that depends on the energy scale as suggested by Markov and Mukhanov (MM) in \cite{Markov:1985py}.
    \item[(ii)] The precise form of $G$ as a function of the energy density $\varepsilon$ can be obtained from the running of $G$ in Asymptotic Safety under a physically motivated cut-off identification.
    \item[(iii)]  The running of $G$ alone captures the qualitative physical features of the model, i.e. we do not explicitly include possible effects from the running of higher dimensional operators, cosmological constant, or matter fields.
\end{itemize}
We do not wish to add any additional matter fields besides a standard one such as dust, radiation or a stiff fluid and the accelerated expansion phase in the early universe, namely inflation, shall be recovered from the matter-gravity coupling. Of course, at low energies we must retrieve the standard field equations of GR. 
With the above assumptions the field equations can be derived in the MM prescription from the following action 
\begin{equation}\label{action}
    S = \int d^4x \sqrt{-g} \left( \frac{R}{8\pi G_N} + 2\chi (\varepsilon) \mathcal{L}_{\rm m} \right),
\end{equation}
where $g$ is the determinant of the spacetime metric $ g_{\mu\nu} $, $ \mathcal{L}_{\rm m} $ is the matter Lagrangian and the matter-gravity coupling depends only on a scalar function $ \chi(\varepsilon) $ of the matter-energy density $\varepsilon$ \cite{Markov:1985py}. Variation of the matter part of the action, with respect to the metric leads to 
\begin{equation}
    \frac{1}{\sqrt{-g}}\delta(2\sqrt{-g}\chi \varepsilon) = 2(\chi \varepsilon),_\varepsilon \delta\varepsilon - \chi \varepsilon g_{\mu\nu} \delta g^{\mu\nu},
\end{equation}
and the variation of the full action leads to the modified Einstein equations
\begin{equation}\label{eq-ee}
    R_{\mu\nu} - \frac{1}{2}g_{\mu\nu}R = 
   8\pi G_N \tilde{T}_{\mu\nu},
\end{equation}
where
\begin{equation}\label{eq-eff-emt}
    \tilde{T}_{\mu\nu} = \left( \varepsilon\chi \right)_{,\varepsilon}T_{\mu\nu} + (\varepsilon^2 \chi_{,\varepsilon}) g_{\mu\nu},
\end{equation}
is the effective energy-momentum tensor while 
\begin{equation}\label{Tmunu}
    T_{\mu\nu} = \left( \varepsilon + P(\varepsilon) \right)u_\mu u_\nu + P(\varepsilon) g_{\mu\nu},
\end{equation}
is the energy-momentum tensor for a ``classical'' perfect fluid, where $P$ is the fluid's pressure which we assume may be obtained from an equation of state $P=P(\varepsilon)$. 
Looking at Eq.~\eqref{eq-ee} and\eqref{eq-eff-emt} we can identify the terms

\begin{equation}
G(\varepsilon) = G_N(\chi \varepsilon)_{,\varepsilon} \; \; \text{and} \; \; \Lambda(\varepsilon) = - 8\pi G_N \varepsilon^2 \chi_{,\varepsilon} 
\end{equation}
as a running gravitational constant $G(\varepsilon)$ and a running cosmological constant $\Lambda(\varepsilon)$. In order to retrieve GR at low densities, for $\chi (\varepsilon)$ we must ensure that as $ \varepsilon \rightarrow 0$ we get $G(\varepsilon) \rightarrow  G_N $, which implies $\chi\rightarrow 1$.

Consider a matter fluid of proper density $ \varepsilon $, 4-velocity $ u^\mu $ with $ u^\mu u_\mu = -1 $ and rest mass density $ \rho $. Mass continuity implies 
\begin{equation}
    (\rho u^\mu)_{;\mu} = 0,
\end{equation}
and for a non-dissipative fluid, we have
\begin{equation}
    \frac{\delta \rho}{\rho} = \frac{\delta\varepsilon}{P(\varepsilon)+\varepsilon}.
\end{equation}

The effective stress-energy tensor needs to be conserved, i.e.
\begin{equation}
    \nabla_\mu \Tilde{T}^\mu_\nu = 0.
\end{equation}
To see how the effective energy density evolves, we can project it along the four-velocity field $u^\mu$ which gives
\begin{equation}\label{tilde-cons}
    \nabla_\mu \left(\Tilde{\varepsilon}u^\mu \right) + \Tilde{P}\nabla_\mu u^\mu = 0,
\end{equation}
where the quantities with \textit{tilde} are the effective fluid's energy-density and pressure that can be obtained from Eq.~\eqref{eq-eff-emt} as 
\bea \label{epsilon-tilde}
\tilde{\varepsilon}&=& (\chi \varepsilon)_{,\varepsilon}\varepsilon  -\varepsilon^2 \chi_{,\varepsilon} =\chi(\varepsilon) \varepsilon, \\ \label{p-tilde}
\tilde{P}&=& (\chi \varepsilon)_{,\varepsilon}P+\varepsilon^2 \chi_{,\varepsilon}.
\eea

Now we can examine the weak and strong energy conditions for the effective fluid. The weak energy condition (w.e.c.) for the perfect fluid in \eqref{Tmunu} holds if $\varepsilon+P\geq0$ and $\varepsilon\geq0$. Then for the effective quantities the w.e.c. is
\begin{equation}
    \tilde{\varepsilon}+\tilde{P}>0, \ \ \ \tilde{\epsilon}\geq0,
\end{equation}
which can be written as
\begin{equation}
    (\varepsilon\chi)_{,\varepsilon}\left(\varepsilon+P\right)>0,\ \ \ \varepsilon\chi\geq0.
\end{equation}
Notice that the behavior of $\chi$ determines the validity of the w.e.c. for the effective fluid. In particular if the w.e.c. holds for the fluid and $\chi>0$ we always have that $\tilde{\varepsilon}\geq 0$, while $\tilde{\varepsilon}+\tilde{P}$ may become negative if $(\varepsilon\chi)_{,\varepsilon}<0$.
A similar reasoning holds for the strong energy condition (s.e.c.) which for the perfect fluid in \eqref{Tmunu} corresponds to $\varepsilon+\sum_{i}P_{i}\geq0$ and $\varepsilon+P_{i}\geq0$ (with $i=1,2,3$ being the spatial directions). The s.e.c. for the effective fluid is then given as 
\begin{equation}
    \tilde{\varepsilon}+3\tilde{P}\geq0, \ \ \ \tilde{\varepsilon}+\tilde{P}\geq0,
\end{equation}
which can be written in terms of the classical fluid components as
\begin{equation}\label{sec}
    (\varepsilon\chi)_{,\varepsilon} \left(\varepsilon+3P\right)+2{\varepsilon}^2\chi_{,\varepsilon}\geq0, \ \ \ (\varepsilon\chi)_{,\varepsilon} \left(\varepsilon+P\right)\geq0.
\end{equation}
Notice that the condition $\tilde{\varepsilon}+3\tilde{P}\geq 0$ must be violated in order to have a phase of accelerated expansion. Then the first of Eq.~\eqref{sec} provides a condition on $\chi$ for which the s.e.c. holds for the classical fluid but is violated for the effective fluid.

The above framework provides a scenario for a variable cosmological constant that naturally emerges from the matter-gravity coupling. The only available freedom is in the choice of one of the two functions $\chi(\varepsilon)$ and $G(\varepsilon)$. 
In the following, we aim to build a model for the early universe for which the gravitational coupling $G(\varepsilon)$ is prescribed from the Asymptotic Safety (AS) program and therefore we will fix the function $G(\varepsilon)$ from the AS paradigm and obtain $\chi (\varepsilon)$ and $\Lambda (\varepsilon)$ accordingly.

If gravity is renormalizable around a non-Gaussian fixed point, one would expect that $ G \sim 1/\kappa^2 $ at very high energies, with $\kappa$ being the characteristic scale of energy at which the physics is being probed \cite{Reuter:1996cp,Weinberg:1980gg,Eichhorn:2018yfc,Pawlowski:2020qer,Reichert:2020mja,Reuter:2019byg,Percacci:2017fkn,Bonanno:2019ilz,Bonanno:2021squ,Bonanno:2020bil}. 
Then 
the variable Newton coupling $G(\kappa)$ can be taken as a function of this characteristic energy scale 
as \cite{Bonanno:2021squ}
\begin{equation}
    G(\kappa)=\frac{G_N}{1+\kappa^2/g_*},
\end{equation}
where the scale $\kappa$ is model dependent 
while $g_*=540\pi/833$ is the UV fixed point. To connect the cutoff scale $\kappa$ and the energy density $\varepsilon$, we follow the prescription discussed in \cite{Bonanno:2019ilz, Platania:2019kyx, Bonanno:2023rzk}. 
Thus we consider the simplest relation between $\kappa$ and  $\varepsilon$ involving both Newton's and Planck constants which can be obtained from dimensional analysis. This is
\be 
\kappa^2=\frac{\hbar G_N}{c^4}\varepsilon,
\ee 
from which we obtain the following form for $G(\varepsilon)$:
\begin{equation}\label{G}
    G(\varepsilon)=\frac{G_N}{1+\frac{\varepsilon}{\varepsilon_c}}, 
\end{equation}
where $\varepsilon_c$ is a dimensional parameter in which we absorb $g_*$ and the other constants. The information on the cutoff scale $\kappa$ is now encoded in $\varepsilon_c$ which we can interpret as a crossover scale separating the Reuter fixed point from the Gaussian fixed point: when $\epsilon/\epsilon_c \gg 1$ the dynamics of the system is essentially determined by the vanishing of the effective coupling constant. In the opposite limit, standard gravity is recovered.  
From Eq.~\eqref{G} we obtain the expressions for $\chi(\varepsilon)$ and $\Lambda(\varepsilon)$ \cite{Bonanno:2023rzk}:
\bea\label{eq-chi-lambda}
        \chi(\varepsilon)&=&\frac{\varepsilon_c}{\varepsilon}\log\left(1+\frac{\varepsilon}{\varepsilon_c}\right), \\ \frac{\Lambda(\varepsilon)}{8\pi G_N} &=&\varepsilon_c\log\left(1+\frac{\varepsilon}{\varepsilon_c}\right)-\frac{\varepsilon}{1+\frac{\varepsilon}{\varepsilon_c}}.
\eea
Notice that for $\varepsilon\rightarrow 0$ we get $\chi\rightarrow 1$ and $\Lambda\rightarrow 0$ thus retrieving the classical GR limit. However for large $\varepsilon$ we see that $\chi\rightarrow 0$ while $\Lambda$ diverges. Looking at Eq.~\eqref{eq-chi-lambda} we can also see that as $\varepsilon$ diverges $\tilde{\varepsilon}=\chi\varepsilon$ diverges as well but more slowly.

\section{Cosmology}\label{cosmology}

The evolution of the universe is governed by the Friedmann equations. We start by considering a Friedmann-Robertson-Walker (FRW) universe with metric
\begin{equation}\label{eq-metric}
    ds^2 = -dt^2 + a^2(t)\left(\frac{dr^2}{1-kr^2} + r^2d\Omega^2\right),
\end{equation}
where $a(t)$ is the scale factor, $k$ is the curvature and $d\Omega^2$ is the line element on the unit sphere. The Friedmann equations for the AS model with MM prescription are then obtained from the metric \eqref{eq-metric} with the effective energy-momentum tensor \eqref{eq-eff-emt} as
\bea \label{eq-FE} 
        H^2 &=& \left(\frac{\dot{a}}{a}\right)^2 = \frac{8\pi G_N}{3}\varepsilon \chi - 
        \frac{k}{a^2} = \frac{8\pi G_N}{3}\tilde{\varepsilon} - \frac{k}{a^2}, \\ \nn
        \frac{\ddot{a}}{a} &=& -\frac{4\pi G_N}{3}\left[ (\varepsilon+3P)\chi + 3\varepsilon\frac{\partial \chi}{\partial \varepsilon}\left( \varepsilon + P \right) \right]=  \\ \label{ddota}
        &=& -\frac{4\pi G_N}{3} (\tilde{\varepsilon}+3\tilde{P}),
\eea
where $ H = \dot{a}/a $ is the Hubble parameter. Then, the evolution of the effective energy density can be easily obtained from Eq.~\eqref{tilde-cons} as
\begin{equation}\label{eq-conservation}
    \dot{\Tilde{\varepsilon}} + 3\frac{\dot{a}}{a}\left(\Tilde{\varepsilon} + \Tilde{P} \right) = \left( \varepsilon \chi \right)_{,\varepsilon}\left[ \dot{\varepsilon} + 3\frac{\dot{a}}{a}\left(\varepsilon + P \right) \right] = 0.
\end{equation}
Since $\left( \varepsilon \chi \right)_{,\varepsilon} \neq 0$ at all times, from the above equation we see that the effective energy momentum is conserved if the original energy momentum is. Namely Eq.~\eqref{tilde-cons} holds if
\begin{equation}
    \dot{\varepsilon} + 3\frac{\dot{a}}{a}\left(\varepsilon + P \right) = 0.
\end{equation}

In the limit, $ \varepsilon \rightarrow 0 $ (and for $\chi=1$), both Friedmann equations of our model reduce to the standard Friedmann equations of GR
\bea 
        H^2 &=& \left(\frac{\dot{a}}{a}\right)^2 = \frac{8\pi G_N}{3}\varepsilon - \frac{k}{a^2}, \\
        \frac{\ddot{a}}{a} &=& -\frac{4\pi G_N}{3} (\varepsilon+3P).
\eea
On the other hand, at high energies, i.e. for $ \varepsilon \rightarrow \infty $, the function $ \chi (\varepsilon) $ goes to zero in such a way that we obtain a de Sitter-like initial phase for the early universe.

In the following, we will consider a matter content given by a perfect fluid described by a linear barotropic equation of state (e.o.s.) $P=w \varepsilon$, and we will pay particular attention to the cases of non relativistic matter (including both baryonic and dark matter) for which $w=0$, radiation for which $w=1/3$ and a stiff fluid for which $w=1$. The case for a stiff equation of state in the early universe was originally proposed by Zel'dovich \cite{Zeldovich:1961sbr, Zeldovich:1972zz} in order to describe a big bang scenario for a universe filled with hot baryons. More recently the same idea has been reevaluated in the context of a universe with a polytropic `dark fluid' where the internal energy plays the role of a stiff fluid and dominates at early times \cite{,Chavanis:2014lra}. 
Interestingly, in our model, due to the vanishing of the gravitation coupling $G(\varepsilon)$, the qualitative behavior of the cosmological model at high densities does not depend on the specific fluid model, with the e.o.s. being important in the determination of the energy scales at which the AS corrections become relevant.

\begin{figure*}[t]
    \begin{center}
        \includegraphics[width=8.6cm]{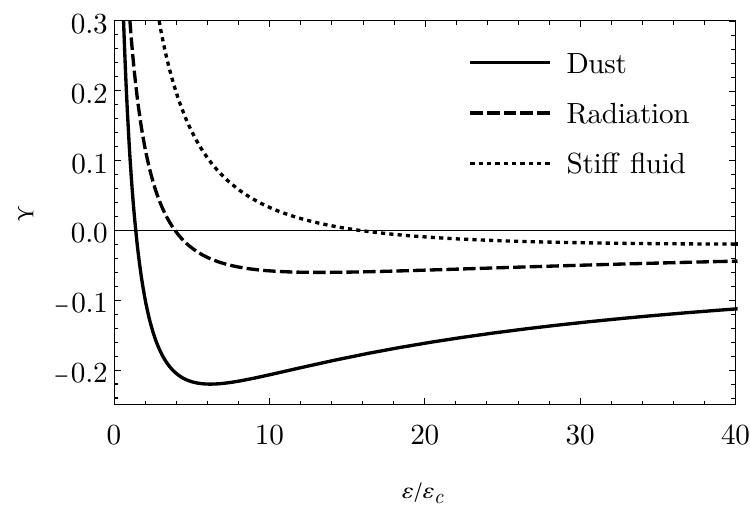}
        \includegraphics[width=8.5cm]{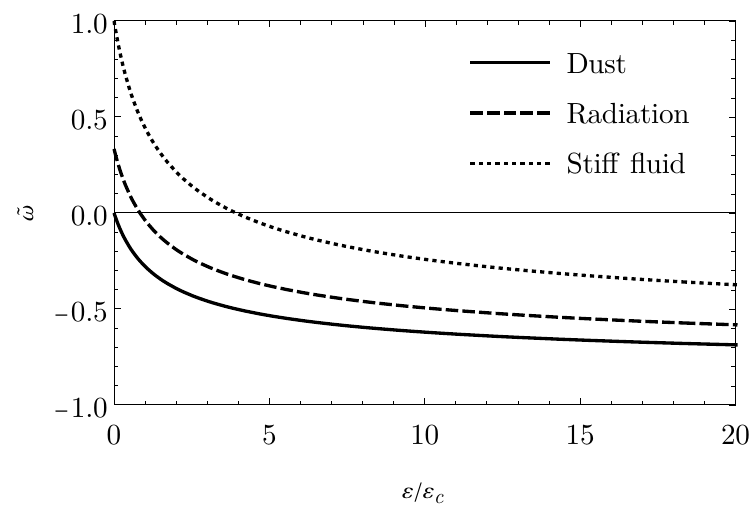}
    \end{center}
    \caption{Left: Dependence of condition for accelerated expansion ($\Upsilon<0$) on the energy density for the single fluid models of dust ($w=0$), radiation ($w=1/3$) and stiff fluid ($w=1$) in AS with MM coupling. Right: Behavior of the effective equation of state $\Tilde{w}$ for the three single fluid models of dust, radiation and stiff fluid in AS with MM coupling as a function of $\varepsilon/\varepsilon_c$. \label{fig-c4in}}
\end{figure*}

We can now take the universe age today as $t_0$ and set $a(t_0)=1$. In the left panel of Fig.~\ref{fig-loglog}, we show the effective energy density $\tilde{\varepsilon}_s = \varepsilon_s \chi$ for a single fluid (denoted by the subscript $s$) in the three cases mentioned above (i.e. $w_s=0,1/3,1$) as a function of the scale factor $a(t)$. The classical energy density $\varepsilon_s$ is obtained from integration of the conservation equation as $ \varepsilon_s \sim 1/a^{3(1+w_s)} $ and $\chi$ is given by Eq.~\eqref{eq-chi-lambda}.
Then for each fluid component of the universe, we define $\Omega_s=\varepsilon_s/
\varepsilon_k$ with $\varepsilon_k$ being the energy density for which the universe is flat, namely $\varepsilon_{k} = 3H^2/8\pi G_N$. Then in terms of the e.o.s. parameter $w_s$ we can write the first Friedmann equation in the following way
\bea\label{eq-fe-oms1}
    \frac{H^2}{H_0^2} &=& \chi \sum_s \Omega_s - \frac{k}{a^2H_0^2},
\eea
with
\bea\label{eq-fe-oms2}
    \Omega_s &=& \frac{\Omega_{s0}}{a^{3(1+w_s)}}, \\ \label{eq-fe-oms3}
\chi &=& \frac{\varepsilon_c}{\varepsilon_{k0}}\frac{\ln\left(1+\frac{\varepsilon_{k0}}{\varepsilon_c} \sum_s \Omega_s \right)}{ \sum_s \Omega_s},
\eea
where $\Omega_{s0}$ are the current values of the density parameters of the respective fluids and $\varepsilon_{k0} = 3H_0^2/8\pi G_N$ is the energy density for the universe to be flat today. To compare the behavior of the scale factor in the $\Lambda$CDM and AS models, we solve the Friedmann Eq.~\eqref{eq-fe-oms1} numerically assuming the fiducial cosmological parameters from \cite{Planck:2018vyg} to find $a(t)$ with the initial condition $a(t_0)=1$. The resulting scale factor is shown in the right panel of Figure \ref{fig-loglog} with a ratio of $\varepsilon_{k0}/\varepsilon_c$ fixed arbitrarily at $10^{-2}$ for illustrative purposes. We see that at late times the model follows the standard $\Lambda$CDM behavior, while at early times the scale factor never reaches zero and asymptotically decreases to $a\rightarrow 0$ with $\dot{a}\rightarrow 0$, thus showing that the big bang singularity is avoided and there is an infinite amount of time in the past for the universe to reach thermal equilibrium.

We are interested in the behavior of $a$ at high energies, i.e. $ \varepsilon \rightarrow \infty $, with the aim of replacing inflation driven by a scalar field with an exponential expansion driven by the variable $G$ in AS. This can be accomplished since the model automatically leads to a quasi-de Sitter phase for $t\rightarrow -\infty$ and no singularity is present at any finite co-moving time. 
The condition for accelerated expansion $ \ddot{a} > 0 $ from Eq.~\eqref{ddota} implies $(\tilde{\varepsilon}+3\tilde{P}) < 0$ which, according to Eq.~\eqref{sec}, 
for a universe dominated by a single fluid with the equation of state, $P=w \varepsilon$ becomes
\be
         \Upsilon = (1+3w)\chi + 3(1+w)\varepsilon \frac{d\chi}{d\varepsilon} < 0. 
\ee
We plot this condition as a function of the energy density in the left panel of Fig.~\ref{fig-c4in} for dust, radiation and stiff fluid. We can clearly see from the figure that $\Upsilon<0$ is satisfied as $\varepsilon\rightarrow\infty$ for all fluid models. We also see that the qualitative behavior for large densities is the same regardless of the fluid model and the e.o.s. plays a role only in determining the energy scale at which $\Upsilon$ becomes negative.
To understand this better we can look at how the effective equation of state relating $\tilde{\varepsilon}$ to $\tilde{P}$ behaves in the early universe. The effective equation of state parameter $\tilde{w}$ is defined as
\begin{equation}
    \Tilde{w}(\varepsilon) = \frac{\Tilde{P}}{\Tilde{\varepsilon}} = \frac{\left(\varepsilon\chi\right)_{,\varepsilon}P+\varepsilon^2 \chi_{,\varepsilon}}{\varepsilon\chi}=w+(1+w)\frac{\varepsilon \chi_{,\varepsilon}}{\chi}.
\end{equation}
If $\varepsilon \chi_{,\varepsilon}/\chi\rightarrow -1$, as is the case for Eq.~\eqref{eq-chi-lambda}, then for large $\varepsilon$ we see that the effective equation of state behaves as
\be\label{limit_w}
    \lim_{\varepsilon\to\infty} \Tilde{w} \rightarrow -1, 
\ee
for any value of $w$. This proves that at high energies, i.e. in the very early universe, the effective equation of state parameter $\Tilde{w}$ shows a limiting de Sitter behavior regardless of the value of $w$.
We show the parameter $\tilde{w}$ in the right panel of Fig.~\ref{fig-c4in} for the cases of dust ($w = 0$), radiation ($w = 1/3$) and stiff fluid ($w = 1$). From the figure, we can also see that at low energies the fluids behave as ordinary perfect fluids. 

\begin{figure*}[t]
    \begin{center}
        \includegraphics[width=14cm]{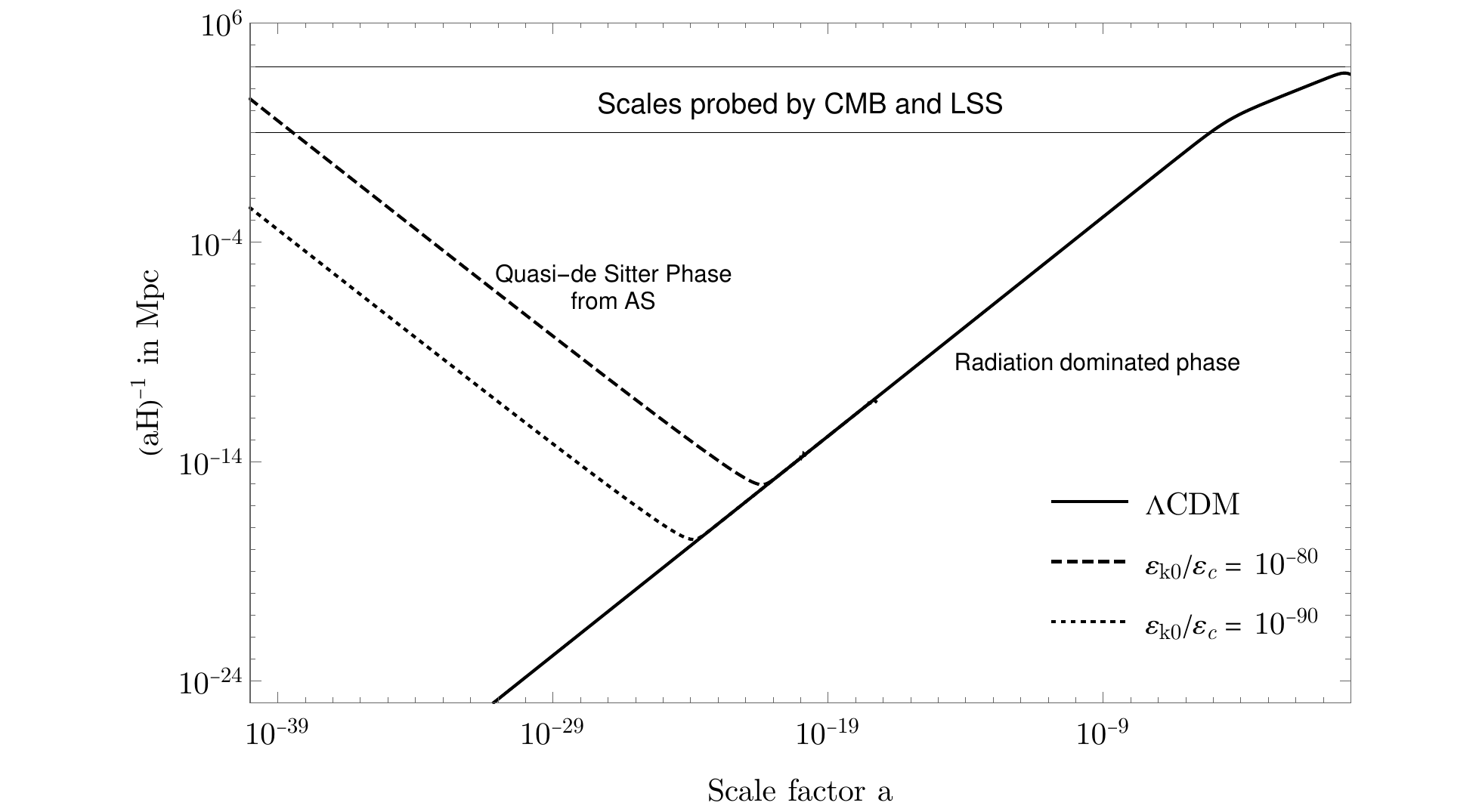}
    \end{center}
    \caption{The comoving Hubble radius as a function of the scale factor $a$. In the plot, the solid line corresponds to the comoving Hubble radius of the $\Lambda$CDM model. The dotted and dashed lines represent the comoving Hubble radius in the AS model for different values of $ \varepsilon_{k0}/\varepsilon_c $ (see Eqs.~\eqref{eq-fe-oms1}-\eqref{eq-fe-oms3}). All other parameters are set from the fiducial values obtained in \cite{Planck:2018vyg}. Notice that depending on the value of $\varepsilon_c$ there is an epoch in the early universe, with corresponding scale factor $a_c$ at which $(aH)^{-1}$ transitions from increasing to decreasing. \label{fig-chr}}
\end{figure*}
 
To see how the cosmological model in AS solves the flatness problem, we can use the Friedmann equations \eqref{eq-FE} and \eqref{ddota} to write
\be \label{eq-domega}
\frac{d\tilde{\Omega}}{d\ln a}=(1+3\tilde{w})(\tilde{\Omega}-1)\tilde{\Omega},
\ee
where $\tilde{\Omega}=\chi\sum_s\Omega_s$ is the effective density parameter. We can clearly see that $\tilde{\Omega} = 1$ is a stable solution since $1+3\tilde{w}<0$ as $\varepsilon \rightarrow \infty$ in AS. This can also be seen easily by introducing small perturbations to the density parameter of the form $\tilde{\Omega} = 1 \pm \delta(a) $ with $ \delta \ll 1 $. At the linear order, Eq.~\eqref{eq-domega} becomes
\be
\frac{d\delta(a)}{d\ln a}=(1+3\tilde{w})\delta(a).
\ee
Solving this differential equation in a regime where $\tilde{w}\simeq {\rm const}.$ yields the following result,
\be\label{eq-delta}
\delta(a)=\delta_i \left(\frac{a}{a_i}\right)^{(1+3\tilde{w})}.
\ee 
Again, since $1+3\tilde{w}<0$ during the AS phase in the early universe, the perturbations decay and the flat solution is stable.

To show that this model also solves the horizon problem we compare the evolution of the comoving Hubble radius $(aH)^{-1}$ in the AS cosmological model with that of the standard $\Lambda$CDM model. In the standard $\Lambda$CDM model without an inflationary phase, the amount of conformal time between the initial singularity and the formation of the CMB is much smaller than the conformal age of the universe today. This leads to patches of angular size $ \theta \sim 2^{\circ} $ on the all-sky projection of the CMB to be causally disconnected. However, the CMB appears to have a uniform temperature map, thus suggesting that those same patches were in causal contact before the emission of the CMB. Adding a phase of exponential expansion to the $\Lambda$CDM model before the hot Big Bang solves the problem. To see this we just need to notice that a decreasing comoving Hubble radius during such a phase, corresponding to increasing the amount of conformal time in the early universe, allows for the past light cones of two points on the CMB to be in causal contact in their past (see Fig.~\ref{fig-chr}).

In the AS model, this quasi-de Sitter phase is achieved automatically via the coupling of $G$ to $\Lambda$. In Fig.~\ref{fig-chr} we show the evolution of the comoving Hubble radius $(aH)^{-1}$ with the Hubble parameter given by the first Friedmann equation \eqref{eq-fe-oms1}. 
It can be clearly seen that in the early universe, during the quasi-de Sitter phase, the comoving Hubble radius was decreasing. The size of the universe at the time of the transition from the exponential expansion to a universe dominated by matter (such as radiation as considered in the figure) depends on the ratio between the current energy density ($\varepsilon_{k0}$) of the universe and the critical, model dependent, density of AS, namely $\varepsilon_c$. The transition from radiation-dominated to the matter-dominated epoch instead occurs around the recombination era and it follows the standard $\Lambda$CDM behavior, regardless of the value of $\varepsilon_c$ (as can be seen from the inflexion of the solid line at the top right of Fig.~\ref{fig-chr}). For comparison, in the same figure, we also show the comoving Hubble radius for the standard $\Lambda$CDM model.

\section{Modeling the early universe}\label{earlyU}

\begin{figure}
    \begin{center}
        \includegraphics[width=8.5cm]{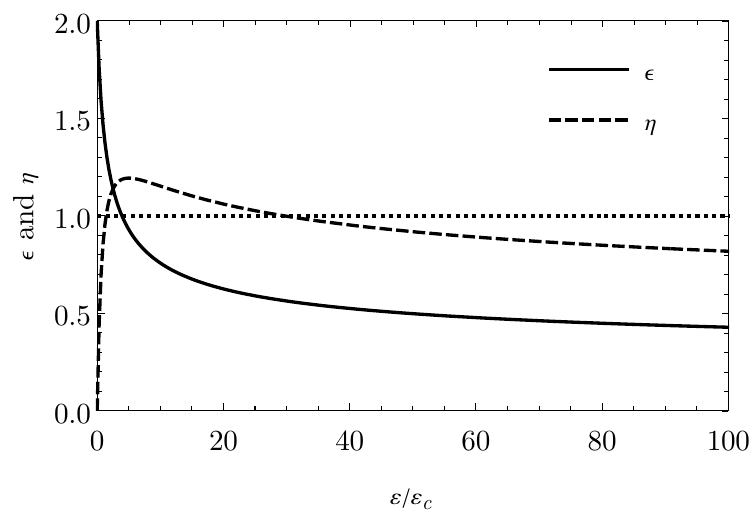}
    \end{center}
    \caption{Left: Behavior of the Hubble slow-roll parameters with respect to the energy density. We have used $P = \varepsilon/3$ in Eq.~\eqref{eq-hub1} and \eqref{eq-hub2} to plot this figure. \label{fig-eos}}
\end{figure}

In order to compare the predictions of the AS model with those of a standard inflationary scenario we need to look at the Hubble slow-roll parameters which indicate the conditions for a sustained inflationary period in the early universe. The Hubble slow-roll parameters are given by \cite{Baumann:2022mni,Lyth:1998xn}
\begin{equation}
        \epsilon \equiv - \frac{\dot{H}}{H^2} = -\frac{d \ln H}{dN}, \ \ \ \ \ 
        \eta \equiv \frac{\dot{\epsilon}}{\epsilon H} = \frac{d\ln \epsilon}{dN},
\end{equation}
where $ dN = Hdt $ refers to the number of e-folds during inflation. The first condition is $ \epsilon < 1 $ and it states that the fractional change of the Hubble parameter per e-fold has to be small. The second condition is $ |\eta| < 1 $, which on the other hand requires $ \epsilon $ to remain small over a large enough number of Hubble times \cite{Baumann:2022mni}. For AS cosmology we get
\bea \label{eq-hub1}
\epsilon&=& \frac{3(1+w)}{2} \frac{\left(\varepsilon\chi\right)_{,\varepsilon}}{\chi}=\frac{3}{2}(1+\tilde{w}), \\ \label{eq-hub2} 
\eta&=& 
-3(1+w)\varepsilon\left[ \frac{\left(\varepsilon \chi\right)_{,\varepsilon\varepsilon}}{\left(\varepsilon \chi\right)_{,\varepsilon}} - \frac{\chi_{,\varepsilon}}{\chi} \right]=\\ \nn
&=&3(1+\tilde{w})(1-\chi).
\eea 
The behaviors of $ \epsilon $ and $ \eta $ for the AS cosmological model with only radiation are shown
in Fig.~\ref{fig-eos}. We can see from the plot that at high energy densities, i.e. as we approach the very early universe, these Hubble slow-roll parameters become smaller than unity which signifies a sustained period of slow-roll inflation. Notice that from Eqs.~\eqref{eq-hub1} and \eqref{eq-hub2} we see that for $a\rightarrow 0$, corresponding to $\tilde{w}\rightarrow -1$, we have that $\epsilon$ and $\eta$ both go to zero.

Now we would like to obtain constraints on the critical density parameter $\varepsilon_c $ by using observational bounds from the primordial power spectrum generated by curvature perturbations during the inflationary period. This analysis will not only enable us to estimate some realistic values of the
characteristic energy scale, but will also allow us to test the observational viability of the AS program of quantum gravity.

\subsection{Linear perturbations and the power spectrum}

To calculate the power spectrum of curvature perturbations generated during the inflationary epoch, we first need to obtain the Lagrangian up to second order in the perturbation variables and then we shall use the method of canonical quantization. To find the second order Lagrangian, we closely follow \cite{Chen:2013kta} where the authors consider a barotropic perfect fluid coupled to gravity and construct a model of inflation which shows a `ultra slow-roll' nature (hence, curvature perturbations are not conserved in the super-Hubble regime). Note that, the effective matter Lagrangian of our model $\mathcal{\tilde{L}_{\rm m}} = \tilde{\varepsilon} = \chi(\varepsilon)\varepsilon$ also behaves as a perfect fluid with a barotropic equation of state. However, in contrast to \cite{Chen:2013kta}, the AS model of inflation shows a slow-roll behavior that closely mimics the standard scalar-field inflationary scenario, i.e., $ \epsilon, \eta \rightarrow 0 $ as $ \varepsilon \rightarrow \infty $. Therefore we expect that the curvature perturbations in our model will be conserved in the super-Hubble scales and the behavior will be similar to that of slow-roll produced by a scalar field. 

We start with the fluid-gravity Lagrangian which can be written in the following way 
\begin{equation}
        \frac{\mathcal{L}}{\sqrt{-g}} = \frac{R}{16\pi G_N} +\mathcal{\tilde{L}}_{\rm m},
\end{equation}
with the matter part given by 
\be \label{42}
\mathcal{\tilde{L}}_{\rm m} = -\tilde{\rho}(1+e) + \lambda_1 (g_{\mu\nu} u^\mu u^\nu + 1) + \lambda_2 (\tilde{\rho}u^\mu)_{;\mu},
\ee 
where $\tilde{\rho}=\tilde{\varepsilon}/(1+e)$ is the rest mass density, $e$ is the specific internal energy, $u^\mu$ is the 4-velocity and $ \lambda_1 $ and $ \lambda_2 $ are Lagrange multipliers for the two constraints: the normalization of the 4-velocity and the conservation of the rest mass density (see \cite{Chen:2013kta} and references therein). 
The variation of the action \eqref{42} with respect to the Lagrange multipliers gives the two constraint equations while the variation with respect to the metric gives the Markov-Mukhanov field equations. At first glance one may think that due to the presence of the additional terms in $\lambda_1$ and $\lambda_2$ one may not retrieve the same field equations. However, for an effective fluid with constant entropy, as it is our case, it is easy to show that one retrieves the correct field equations (see \cite{Ray:1972} for details).

Now, to construct the quadratic action, we follow the standard procedure of splitting the action in the $ 3+1 $ formalism according to which the metric can be written as
\begin{equation}
    ds^2 = -\alpha^2dt^2 + h_{ij}(dx^i+\beta^idt)(dx^j+\beta^jdt),
\end{equation}
where $\alpha$ is the lapse function and $\beta^i$ is the shift vector. A spatially-flat FRW background corresponds to $\alpha = 1$ and $ \beta^i = 0 $. Putting the above metric into the action leads to the ADM action 
\begin{equation}
    S = \int dt d^3x \sqrt{h}\alpha(\mathcal{L}_{\rm G} + \mathcal{\tilde{L}}_{\rm m}) ,
\end{equation}
where the gravity part $\mathcal{L}_{\rm G}=R/(16\pi G_N)$ is now written as 
\begin{equation}
        \mathcal{L}_{\rm G} = \frac{1}{16\pi G_N}\left[ ^{(3)}R + \alpha^{-2}\left(K_{ij}K^{ij}- K^2\right) \right].
\end{equation}
Here $^{(3)}R$ is the Ricci scalar constructed out of the three-dimensional metric $h_{ij}$ and $K_{ij}$ is the extrinsic curvature on a constant $t$ hypersurface
\begin{equation}
    K_{ij} = \frac{1}{2}\dot{h}_{ij} -^{(3)}\nabla_j \beta_i - ^{(3)}\nabla_i \beta_j,
\end{equation}
with $^{(3)}\nabla_i$ representing the covariant derivative with respect to the three dimensional metric $h_{ij}$ and $K$ being the trace of $K_{ij}$. The variation of the ADM action with respect to $\lambda_1$, $\lambda_2$, the lapse and shift (which are Lagrange multipliers of the gravitational system) leads to four constraint equations. These constraints can be solved for the Lagrange multipliers and later can be used in the action in favor of the perturbation variable of interest, which in our case would be the scalar curvature perturbations.    

Now to introduce linear perturbations to the system we choose the comoving gauge, in which
\begin{equation}
    u^{\mu} = ( -1 + u, 0, 0, 0 ), \ \ \ \ \ \ h_{ij} = a^2(t)e^{2\mathcal{R}}\delta_{ij}.
\end{equation}
Here $u$ denotes the velocity scalar potential to all order in perturbations and $\mathcal{R}$ denotes the curvature perturbations in the comoving gauge. With these perturbations, eliminating the Lagrange multipliers, the second order Lagrangian can now be written as (see \cite{Chen:2013kta} for details)
\begin{equation}
    \mathcal{L}_{\mathcal{R}}^{(2)} = \frac{a^3}{8\pi G_N} \left[ \frac{\epsilon}{\tilde{c}_s^2}\dot{{\mathcal{R}}}^2 - \frac{\epsilon}{a^2}(\partial\mathcal{R})^2 \right],
\end{equation}
where $\tilde{c}_s^2 =\delta\tilde{P}/\delta\tilde{\varepsilon} $ is the square of the propagation speed of the pressure perturbations of the effective fluid in the rest frame of the fluid. Notice that from $\tilde{P} = \tilde{P}(\tilde{\varepsilon})$ follows that $\delta\tilde{P}/\delta\tilde{\varepsilon} = \dot{\tilde{P}}/\dot{\tilde{\varepsilon}}$. We can write this explicitly in terms of the classical fluid variables and the coupling function $\chi$ as
\begin{align}\label{eq-cs1}
    \tilde{c}_s^2  = \frac{\dot{P}}{\dot{\varepsilon}} + (w+1)\varepsilon\frac{(\chi\varepsilon)_{,\varepsilon\varepsilon}}{(\chi\varepsilon)_{,\varepsilon}}. 
\end{align}
However, the fact that the spatial part of the Bianchi identity for the classical fluid is not satisfied, i.e. $\nabla_\mu T^\mu_i \neq 0$, indicates the presence of non-adiabatic pressure perturbations \cite{Gordon:2004ez}. As a consequence, for the classical fluid $\dot{P}/\dot{\varepsilon} \neq \delta P/\delta \varepsilon $. From the explicit evaluation of the Bianchi identity for the effective fluid, we obtain the following relation for the classical fluid
\begin{align}\label{eq-cs2}
    \frac{\dot{P}}{\dot{\varepsilon}} = \frac{\delta P}{\delta \varepsilon} - (w+1)\varepsilon\frac{(\chi\varepsilon)_{,\varepsilon\varepsilon}}{(\chi\varepsilon)_{,\varepsilon}}.
\end{align}
Now from Eq.~\eqref{eq-cs1} and \eqref{eq-cs2} we get,
\begin{align}
    \tilde{c}_s^2 = \frac{\delta P}{\delta \varepsilon} = w.
\end{align}
In the following, in order to simplify the notation, we set $8\pi G_N = 1$. By defining $z^2 = 2a^2\epsilon/\tilde{c}_s^2$ the second order action can be rewritten as 
\begin{equation}\label{action-z}
     S^{(2)} = \frac{1}{2} \int d\tau d^3x z^2 \left[ (\mathcal{R}')^2 - \tilde{c}_s^2 (\partial\mathcal{R})^2 \right],  
\end{equation}
where the primed quantity denotes the derivative with respect to the conformal time $\tau$ which is defined as $\tau = \int dt/a $. For convenience, we introduce a change of variable from $z$ to $v$ given by $v = z\mathcal{R}$, where $v$ is known as the Mukhanov-Sasaki variable. In terms of the new variable, the action \eqref{action-z} becomes
\begin{equation}\label{eq-action-ms}
    S^{(2)} = \frac{1}{2} \int d\tau d^3x  \left[ v'^2 - \tilde{c}_s^2 (\partial v)^2 - m(\tau)^2 v^2 \right],
\end{equation}
where 
\begin{eqnarray} \nonumber
 m(\tau)^2 &=& -\frac{z''}{z} = (aH)^2 \left[2 - \epsilon + \frac{3\eta}{2}  -\frac{\dot{\tilde{c}}_s}{\tilde{c}_s H} - \left(\frac{\eta}{2}\right)^2 +\right. \\ 
 &+&\left. \frac{\ddot{\epsilon}}{2\epsilon H} - \frac{\dot{\tilde{c}}_s \eta}{\tilde{c}_s H}  - \frac{\ddot{\tilde{c}}_s}{\tilde{c}_s H^2} + 2\left( 
\frac{\dot{\tilde{c}}_s}{\tilde{c}_s H} \right)^2 \right] ,
\end{eqnarray} 
acts like a time dependent mass parameter. In the quasi-de Sitter era, we have 
\begin{equation}
    aH \simeq -\frac{1}{\tau} (1+\epsilon).
\end{equation}
Therefore, we can simplify the mass parameter as
\begin{equation}
    m(\tau)^2 \simeq \frac{2+3\epsilon + 3\eta/2}{\tau^2},
\end{equation}
where we have only kept the terms linear in $\epsilon$ and $\eta$. Variation of the action leads to the equation of motion for the Mukhanov-Sasaki variable $v$,
\begin{equation}
    v'' + \tilde{c}_s^2 \nabla^2 v + m(\tau)^2 v = 0. 
\end{equation}
This equation is known in the literature as the Mukhanov-Sasaki equation \cite{Sasaki:1986hm,Mukhanov:1988jd}. 

Since the action \eqref{eq-action-ms} is in the canonically normalized form, in order to quantize the system, we shall apply the method of canonical quantization. Working in the Heisenberg picture, we promote $v$ as operator $ \hat{v} $ and expand it in terms of creation of annihilation operators
\bea
        \hat{v}(\tau,\vec{x}) &=& \int \frac{d^3\vec{k}}{2\pi^3}\hat{v}_k(\tau) e^{i\vec{k}\cdot\vec{x}} = \\ \nn
        &=& \int \frac{d^3\vec{k}}{2\pi^3} \left[ v_k(\tau) e^{i\vec{k}\cdot\vec{x}} \hat{a}_{\vec{k}} + v_k^*(\tau) e^{-i\vec{k}\cdot\vec{x}} \hat{a}_{\vec{k}}^\dagger  \right],
\eea
where the Fourier modes $\hat{v}_k$ are those of the one-dimensional time-dependent quantum harmonic oscillators, that can be written as
\begin{equation}
    \hat{v}_k(\tau) = v_k(\tau) \hat{a}_{\vec{k}} + v_k^* (\tau) \hat{a}_{-\vec{k}}^\dagger,
\end{equation}
with $\hat{a}_{\vec{k}}$ and $\hat{a}_{-\vec{k}}^\dagger$ being the annihilation and creation operators, respectively, which satisfy the usual commutation relations
\begin{equation}
    \left[ \hat{a}_{\vec{k}_1}, \hat{a}_{\vec{k}_2}^\dagger \right] = (2\pi)^3\delta^3_D(\vec{k}_1 - \vec{k}_2), 
\end{equation}
and the vacuum state is defined by $\hat{a}_{\vec{k}}|0\rangle = 0$.
The mode functions $v_k(\tau)$ satisfy the Mukhanov-Sasaki equation
\begin{equation}\label{eq-ms}
    \begin{aligned}
        v''_k + \left(\tilde{c}_s^2k^2 + m(\tau)^2 \right)v_k = 0,
    \end{aligned}
\end{equation}
and they are canonically normalized to 
\begin{equation}
    v_k v^{*\prime}_{k} - v_k^{\prime} v_k^* = i\hbar,
\end{equation}
so that, now the quantum field $\hat{v}$ and its conjugate momentum $ \hat{\Pi}_v = \hat{v}'$ satisfy the usual canonical commutation relation 
\begin{equation}
    \left[ \hat{v}(\tau, \vec{x}_1), \hat{\Pi}_v (\tau, \vec{x}_2) \right] = i\hbar \delta_D^3 (\vec{x}_1 - \vec{x}_2).
\end{equation}
We can now discuss the quantum statistics of the operator $\hat{v}(\tau)$. The expectation value of $\hat{v}$ vanishes, i.e. $ \langle \hat{v} \rangle \equiv \langle 0 | \hat{v} | 0 \rangle = 0 $. However, the variance of the fluid fluctuations receive non-zero quantum contributions,
\begin{eqnarray}
        \langle |\hat{v}|^2 \rangle &\equiv& \langle 0 | \hat{v}(\tau, \vec{x})\hat{v}(\tau, \vec{x}) | 0 \rangle = \nonumber \\
        &=& \int \frac{d^3k}{(2\pi)^3}\int \frac{d^3k'}{(2\pi)^3} \langle 0 | \big( v_k^*(\tau)\hat{a}^\dagger_{\vec{k}} + v_k(\tau)\hat{a}_{\vec{k}} \big)\times \nonumber \\
        && \ \ \ \ \ \ \ \ \ \ \ \ \ \ \ \ \ \ \   \big( v_{k'}(\tau)\hat{a}_{\vec{k'}} + v_{k'}^*(\tau)\hat{a}^\dagger_{\vec{k'}} \big) | 0 \rangle = \nonumber \\
        &=& \int d \ln k \frac{k^3}{2\pi^2} |v_k(\tau)|^2 =  \int d\ln k \ \mathcal{P}_v(k,\tau),
\end{eqnarray}
where we have defined the dimensionless power spectrum as
\begin{equation}\label{eq-powerspectrum}
    \mathcal{P}_v(k,\tau) \equiv \frac{k^3}{2\pi^2} |v_k(\tau)|^2.
\end{equation}
To find the power spectrum of the quantum fluid fluctuations in our model we need to solve Eq.~\eqref{eq-ms} with the appropriate boundary condition. The Mukhanov-Sasaki equation can be written as a Bessel equation.
Taking
\begin{equation}
    \tau^2m(\tau)^2 = \nu^2 -1/4 ,
\end{equation}
we define a new time variable
\begin{equation}
    T = -\tilde{c}_s k \tau ,
\end{equation}
in terms of which Eq.~\eqref{eq-ms} becomes
\begin{equation}
    \frac{d^2v_k}{dT^2} + \left[ 1 - \frac{1}{T^2}\left(\nu^2 - \frac{1}{4}\right) \right]v_k = 0.
\end{equation}
All modes undergo Hubble exit at $ T/\tilde{c}_s = 1 $, i.e. $k = aH$ with sub (super)-Hubble scales corresponding to $ T/\tilde{c}_s \gg (\ll) 1 $. On the other hand, the modes cross the sound horizon when $aH = \tilde{c}_s k$ or $T=1$.

Finally, we use another redefinition of the variables setting $ F = v_k/\sqrt{T} $ to write the Mukhanov-Sasaki equation in the following form
\begin{equation}\label{eq-F}
    \frac{d^2F}{dT^2} + \frac{1}{T} \frac{dF}{dT} + \left[ 1 - \frac{\nu^2}{T^2} \right]F = 0.
\end{equation}
In this case, the equation can be solved analytically.
The general solution of Eq.~\eqref{eq-F} can be written in terms of Hankel functions as
\begin{equation}
    F(T) = C_1 \mathrm{H}_\nu^{(1)}(T) + C_2 \mathrm{H}_\nu^{(2)}(T),
\end{equation}
with $C_1$  and $C_2$ fixed by the boundary conditions. Hence the solution of the Mukhanov-Sasaki equation is
\begin{equation}
    v_k(T) = \sqrt{T}\left[ C_1 \mathrm{H}_\nu^{(1)}(T) + C_2 \mathrm{H}_\nu^{(2)}(T) \right].
\end{equation}
In the limit $\tilde{c}_sk \ll aH$, the Hankel functions take the form
\bea\label{eq-hankel}
        \mathrm{H}_\nu^{(1)} (T) \Big|_{T \rightarrow 0} &\simeq& \sqrt{\frac{2}{\pi}} e^{-i\frac{\pi}{2}} 2^{\nu - \frac{3}{2}} \frac{\Gamma (\nu)}{\Gamma(\frac{3}{2})} T^{-\nu}, \\ \nn
        \mathrm{H}_\nu^{(2)} (T) \Big|_{T \rightarrow 0} &\simeq& -\sqrt{\frac{2}{\pi}} e^{-i\frac{\pi}{2}} 2^{\nu - \frac{3}{2}} \frac{\Gamma (\nu)}{\Gamma(\frac{3}{2})} T^{-\nu}.
\eea
To find $C_1$ and $C_2$ we use the Bunch-Davies conditions \cite{Bunch:1978yq} for the mode functions deep in the sub-Hubble limit, i.e. $\tilde{c}_sk\gg aH$ (or $T\rightarrow \infty$). Namely
\begin{equation}
    v_k(T) \Big|_{T\rightarrow \infty} \rightarrow \frac{1}{\sqrt{2\tilde{c}_sk}} e^{iT} = \sqrt{T} C_1 \mathrm{H}_\nu^{(1)} (T)\Big|_{T \rightarrow \infty}
\end{equation}
which yields
\begin{equation}
    C_1 = \frac{1}{\sqrt{2\tilde{c}_sk}} \sqrt{\frac{\pi}{2}} e^{i\left( \nu + \frac{1}{2} \right)\frac{\pi}{2}}, \ \ \ \ C_2 = 0. 
\end{equation}
Therefore the final expression for the mode functions becomes
\begin{equation}\label{eq-vk-sol}
v_k(T) =\frac{ e^{i\left( \nu + \frac{1}{2} \right)\frac{\pi}{2}} }{2} \sqrt{\frac{\pi T}{\tilde{c}_sk}}  \mathrm{H}_\nu^{(1)} (T).     
\end{equation}
Finally, going back to Eq.~\eqref{eq-powerspectrum}, we can write the power spectrum of comoving curvature perturbations in our model as
\begin{equation}
    \mathcal{P}_{\mathcal{R}}(k) = \frac{k^3}{2\pi^2}|\mathcal{R}_k|^2 = \frac{k^3}{2\pi^2} \frac{|v_k|^2}{z^2}, 
\end{equation}
where, $v_k$ is given by Eq.~\eqref{eq-vk-sol} and the Hankel function in the super-Hubble limit comes from from Eq.~\eqref{eq-hankel}. With these replacements, the power spectrum becomes
\begin{equation}
    \mathcal{P}_{\mathcal{R}}(k) = 2^{2\nu-3}\left( \frac{\Gamma(\nu)}{\Gamma(\frac{3}{2})} \right)^2 \frac{1}{8\pi^2\epsilon \tilde{c}_s} \left(\frac{H}{m_{\rm pl}}\right)^2 \left( \frac{\tilde{c}_s k}{aH} \right)^{3-2\nu},
\end{equation}
where $a$, $H$ and $\epsilon$ are evaluated at the sound horizon exit and we have reinstated the factor of $ m_{\rm pl}^2 = 1/8\pi G_N$. Now, the spectral index of the scalar perturbations is easily determined as
\begin{equation}\label{eq-ns}
n_s - 1 \simeq 3 - 2\nu \simeq -2\epsilon - \eta.     
\end{equation}
As we can see, since $\{\epsilon, |\eta|\} \ll 1$ in the very early universe, perturbations in the effective fluid produces a nearly scale-invariant power spectrum.
Finally the tensor power spectrum $\mathcal{P}_{T}$ is obtained in exactly the same manner as the single scalar field slow-roll case. We obtain 
\begin{equation}
    \mathcal{P}_{T} = \frac{2H^2}{\pi^2 m_{\rm pl}^2},
\end{equation}
so that the tensor-to-scalar ratio $r$ is given by
\begin{equation}\label{eq-r}
    r \simeq 16\epsilon \tilde{c}_s.
\end{equation}
Notice that $r$ in our model has a factor $\tilde{c}_s$, which makes it distinguishable from standard scalar field slow-roll inflation.

\subsection{Constraints from observations}

\begin{figure*}[t]
    \begin{center}
        \includegraphics[width=5.5cm]{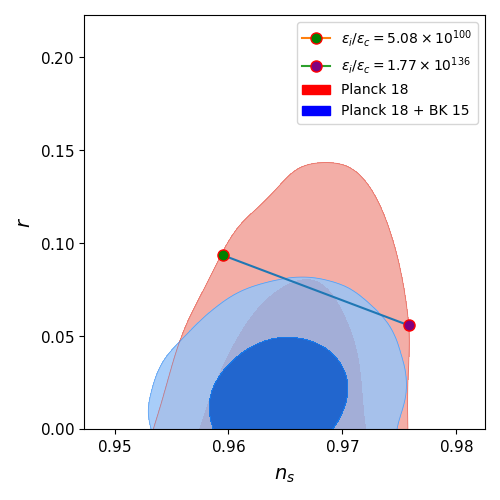}
        \includegraphics[width=5.5cm]{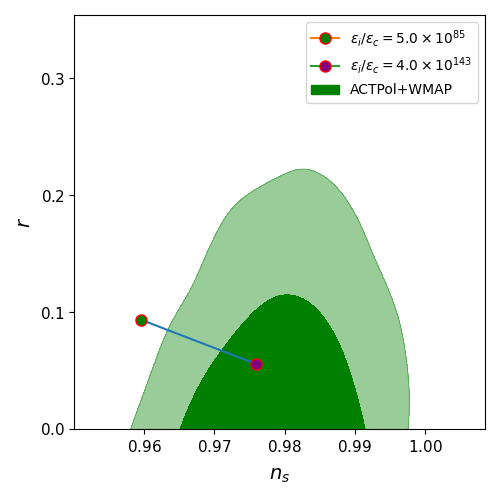}
        \includegraphics[width=5.5cm]{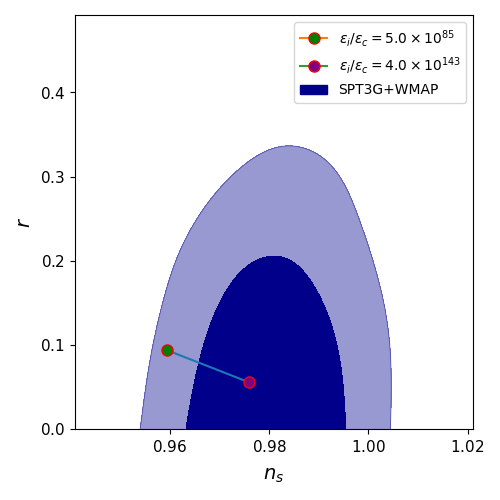}
    \end{center}
    \caption{Constrains on $\varepsilon_i/\varepsilon_c$ for the AS model of inflation in a radiation dominated universe ($w=1/3$) from the marginalized contours of Planck and Planck+BK15 (left panel), ACTPol+WMAP (mid panel) and SPT3G+WMAP (right panel) data. The contours are obtained for $n_s$ and $r$ using marginalized joint 68\% and 95\% confidence-level regions from the observational data.\label{fig-pl-bk}}
\end{figure*}

\begin{table*}[]
\begin{tabular}{|l|l|l|l|l|l|l|l|}
\hline
$w$ &
  ($n_s, r$) &
  $\varepsilon_i/\varepsilon_c$ &
  $\varepsilon_i$ GeV$^4$ &
  $\tilde{\varepsilon}_i $ GeV$^4$ &
  $(\varepsilon_i)^{1/4}$ GeV &
  $(\tilde{\varepsilon}_i)^{1/4}$ GeV &
  $N_{\rm tot}$ \\ \hline
1/3 & (0.959, 0.09) & $5 \times 10^{85}$   & $1.03 \times 10^{162}$   & $4.09 \times 10^{78}$    & $3.2 \times 10^{44}$    & $4.5 \times 10^{19}$   & 48.9 \\ \cline{2-8} 
    & (0.975, 0.055) & $4 \times 10^{143}$   & $8.29 \times 10^{219}$   & $6.85 \times 10^{78}$    & $9.54 \times 10^{54}$   & $5.11 \times 10^{19}$   & 82.3 \\ \hline
1   & (0.965, 0.13) & $  10^{150} $ & $ 2.07 \times 10^{226} $ & $ 7.16 \times 10^{78} $ & $ 3.79 \times 10^{56} $ & $ 5.17 \times 10^{19} $ & 57.1  \\ \cline{2-8} 
    & (0.975, 0.10) & $ 10^{205} $ & $ 2.07 \times 10^{281} $ & $ 9.79 \times 10^{78} $ & $ 2.13 \times 10^{70} $  & $ 5.59 \times 10^{19} $ & 78.21 \\ \hline
\end{tabular}
\caption{Values of the energy density $\varepsilon_i$ and its effective counterpart $\tilde{\varepsilon}_i$ at the epoch of horizon crossing from the fit with the data assuming $\varepsilon_c \sim m_{\rm pl}^4$ for both radiation (i,e, $w=1/3$ as in Fig.~\ref{fig-pl-bk}) and stiff fluid ($w=1$). The last column shows the number of the universe's e-folds during the quasi-de Sitter period.}
\label{tab1}
\end{table*}

To compare the predictions of our model of inflation in AS with current observations we consider data sets obtained from Planck and other experiments. We specifically consider the observational constraints on the scalar spectral index $n_s$ and the tensor-to-scalar ratio $r$ based on the Planck 2018 and BICEP2/Keck-Array 2015 (Planck 18 + BK 15) combined data sets \cite{Planck:2018nkj,Planck:2018vyg,Planck:2019nip,Planck:2018jri,BICEP2:2018kqh}, the Atacama Cosmology Telescope DR4 likelihood combined with the WMAP satellite data (ACTPol+WMAP) \cite{ACT:2020gnv,WMAP:2012nax,Forconi:2021que} and the South-Pole Telescope polarization measurements combined with the WMAP satellite data (SPT3G+WMAP) \cite{SPT-3G:2021eoc,WMAP:2012nax,Forconi:2021que}. Planck temperature, polarization, and lensing data have determined the spectral index of scalar perturbations to be $n_s = 0.9649 \pm 0.0042 $ at $68\%$ confidence level (CL) with no evidence of a scale dependence. On the other hand, the Planck data (Planck TT,TE,EE+lensing+lowEB) puts a $95\%$ CL upper limit on the tensor-to-scalar ratio, $r\leq0.16$ which is further improved by combining with the BICEP2/Keck Array data to obtain $r<0.056$ \cite{Planck:2018vyg,Planck:2018jri}. These are the strongest constraints on $n_s$ and $r$ to date.

Now, we can use the slow roll parameters $\epsilon$ and $\eta$ in our model to evaluate $n_s$ and $r$ from Eq.~\eqref{eq-ns} and \eqref{eq-r} 
in order to put constraints from observation on the value of $\varepsilon_i/\varepsilon_c$ during the inflationary period, where $ 
\varepsilon_i $ is the energy scale of inflation. In view of the fact that for $\varepsilon=\varepsilon_c$ from Eq.~\eqref{G} we get $G(\varepsilon_c)=G_N/2$, we need $\varepsilon_i>>\varepsilon_c$ in order to achieve the de Sitter phase for which $G(\varepsilon_i)\simeq 0$. For example, for a critical density of the order of Planck density, $\varepsilon_c \sim m_{\rm pl}^4$, the energy scale of the universe at the horizon exit $\varepsilon_i$ would have to be above the Planck density, i.e. $\varepsilon_i^{1/4} \sim 10^{44} {\rm GeV}$ corresponding to $\tilde{\varepsilon}_i^{1/4} \sim 10^{19.5} {\rm GeV}$.

In Fig.~\ref{fig-pl-bk} we plot the marginalized contours of $n_s$ and $r$ for the datasets considered along with the predictions of the AS model calculated from Eq.~\eqref{eq-ns} and \eqref{eq-r} for a radiation dominated universe. We can see that our model agrees well with all three datasets. 
In all three plots of Fig.~\ref{fig-pl-bk}, $ \varepsilon_i/\varepsilon_c $ varies along the straight line with the points at the ends taking values $ \varepsilon_i/\varepsilon_c \simeq 5 \times 10^{85}$ and $\varepsilon_i/\varepsilon_c \simeq 4 \times 10^{143} $ (assuming $w = 1/3$). 
Similarly, we can obtain the constraints on the energy scales of inflation for the stiff fluid model as $ \varepsilon_i/\varepsilon_c \simeq 10^{150}$ and $\varepsilon_i/\varepsilon_c \simeq 10^{205} $. These values, together with the corresponding number of e-folds during the inflation period, are shown in Table~\ref{tab1} for both radiation and stiff fluid.

The number of e-folds from the time of horizon exit to the end of the quasi-de Sitter period is given by 
\begin{equation}\label{eq-efolds1}
    N_{tot} = \ln\left(\frac{a_e}{a_i}\right).
\end{equation}
Here $a_i$ and $a_e$ are the values of the scale factor at the moment of horizon exit and the end of the quasi-de Sitter period. The end of accelerated expansion is defined by $\ddot{a}_e=0$.
We then use Eq.~\eqref{eq-conservation} to write
\begin{equation}
    \frac{\varepsilon_e}{\varepsilon_i} = \left( \frac{a_e}{a_i} \right)^{-3(1+w)},
\end{equation}
where $\varepsilon_i$ and $\varepsilon_e$ are the energy densities at the moment of horizon exit and the end of the quasi-de Sitter period. Using the above equation, the number of e-folds becomes
\begin{equation}\label{eq-efolds2}
    N_{tot} = \ln \left( \frac{\varepsilon_i}{\varepsilon_e} \right)^{1/3(1+w)}. 
\end{equation}
To derive the ratio of energy densities at the two epochs, we use the constraints we obtained from CMB observations to find $\varepsilon_i/\varepsilon_c$ and the condition for the end of accelerated expansion to find $\varepsilon_e/\varepsilon_c$. 
First, from the CMB constraints discussed above, we found that the ratio $\varepsilon_i/\varepsilon_c$ that gives $n_s=0.965$, $r=0.08$ for radiation ($w = 1/3$) is
\begin{equation}
    \frac{\varepsilon_i}{\varepsilon_c} \simeq 10^{100} .
\end{equation}
Then from the condition $ \tilde{\varepsilon} + 3 \tilde{P} = 0 $ which holds at the end of accelerated expansion, we get
\begin{equation}
    \frac{\varepsilon_e}{\varepsilon_c} \simeq 3.921.
\end{equation}
These two equations give the ratio of the energy density at the horizon exit and the end of the quasi-de Sitter period
\begin{equation}
    \frac{\varepsilon_i}{\varepsilon_e} \simeq 2.5 \times 10^{99},
\end{equation}
from which we can calculate the number of e-folds using Eq.~\eqref{eq-efolds2}, which gives
\begin{equation}
    N_{tot} \simeq 57.2 \ \ \ {\rm for} \ \ w = 1/3.
\end{equation}
Similarly, we can perform the same calculations for dust and stiff fluid, for which the numbers of e-folds are 
\begin{equation}
    \begin{aligned}
        &N_{tot} \simeq 57.45 \ \ \ {\rm for} \ \ w = 0, \\
        &N_{tot} \simeq 57.1 \ \ \ {\rm for} \ \ w = 1.
    \end{aligned}
\end{equation}
Similar calculations for the ratio $\varepsilon_i/\varepsilon_c$ taking $n_s \simeq 0.975$ and $r < 0.10$ for the three matter models give $N_{tot}\simeq 81$.
The values are consistent with the estimates of $N$ from scalar field inflation thus confirming that the model can be considered a viable alternative to standard inflationary models.

\section{Summary and Discussions}\label{summary}

We considered a modification of the standard cosmological model at high densities that allows for a phase of accelerated expansion without the introduction of additional fields. This was achieved with the introduction of a matter-gravity coupling $\chi$ in the action that leads to an effective theory with varying gravitational coupling $G$ and $\Lambda$ in the field equations. It is worth noting that due to the monotonic behavior of the coupling $\chi$, we can not use the Markov-Mukhanov mechanism to explain the observed late-time acceleration of the universe within a single framework. There exist other approaches to variable $G$ and $\Lambda$ that attempts at explaining the late-time acceleration, see e.g. \cite{deCesare:2016dnp}.
We showed that the choice of $G(\varepsilon)$ consistent with the Asymptotic Safety paradigm allows us to reproduce a phase of accelerated expansion in the early universe without the aid of an ad hoc scalar field.

By treating the model at perturbative level we calculated the power spectrum of scalar curvature fluctuations. Curvature perturbations are seen to be conserved in the super-horizon limit and hence the power-spectrum is nearly scale invariant. We calculated the spectral index and the tensor-to-scalar ratio of the perturbations and used them to put constraints on the model's parameters from the latest observational data. We showed that our results are consistent with the experimental bounds of the existing observations. 
In particular, we showed that there are ranges of values 
for $\varepsilon_i/\varepsilon_c$, with $\varepsilon_i$ and $\varepsilon_c$ being the energy density at the horizon exit and the critical energy density (the model's only free parameter) respectively, for which the observational constraints from the CMB data can be satisfied.  
Finally, we expect that the data obtained from future experiments such as CMB-S4 \cite{CMB-S4:2020lpa,CMB-S4:2023zem}, LiteBird \cite{LiteBIRD:2022cnt}, CORE \cite{CORE:2017oje}, CMB-Bh\={a}rat \cite{Adak:2021lbu}, etc. will be able to test the validity of the model presented here.

\section*{Acknowledgement}
AZ would like to thank Catania Astrophysical Observatory - INAF-  for its warm hospitality during the preparation of the manuscript.
DM, HC and AZ acknowledge support from Nazarbayev University Faculty Development Competitive Research Grant No. 11022021FD2926. 
The authors thank William Giar\`e and his group for giving us access to some of the data products used in this article. The authors also thank Antonio De Felice for useful comments and discussions.

\bibliographystyle{apsrev}
\bibstyle{apsrev}
\bibliography{ref}

\begin{thebibliography}{67}
\expandafter\ifx\csname natexlab\endcsname\relax\def\natexlab#1{#1}\fi
\expandafter\ifx\csname bibnamefont\endcsname\relax
  \def\bibnamefont#1{#1}\fi
\expandafter\ifx\csname bibfnamefont\endcsname\relax
  \def\bibfnamefont#1{#1}\fi
\expandafter\ifx\csname citenamefont\endcsname\relax
  \def\citenamefont#1{#1}\fi
\expandafter\ifx\csname url\endcsname\relax
  \def\url#1{\texttt{#1}}\fi
\expandafter\ifx\csname urlprefix\endcsname\relax\def\urlprefix{URL }\fi
\providecommand{\bibinfo}[2]{#2}
\providecommand{\eprint}[2][]{\url{#2}}

\bibitem[{\citenamefont{Hinshaw et~al.}(2013)}]{WMAP:2012nax}
\bibinfo{author}{\bibfnamefont{G.}~\bibnamefont{Hinshaw}} \bibnamefont{et~al.}
  (\bibinfo{collaboration}{WMAP}), \bibinfo{journal}{Astrophys. J. Suppl.}
  \textbf{\bibinfo{volume}{208}}, \bibinfo{pages}{19} (\bibinfo{year}{2013}),
  \eprint{1212.5226}.

\bibitem[{\citenamefont{Akrami et~al.}(2020)}]{Planck:2018jri}
\bibinfo{author}{\bibfnamefont{Y.}~\bibnamefont{Akrami}} \bibnamefont{et~al.}
  (\bibinfo{collaboration}{Planck}), \bibinfo{journal}{Astron. Astrophys.}
  \textbf{\bibinfo{volume}{641}}, \bibinfo{pages}{A10} (\bibinfo{year}{2020}),
  \eprint{1807.06211}.

\bibitem[{\citenamefont{Aghanim et~al.}(2020{\natexlab{a}})}]{Planck:2018nkj}
\bibinfo{author}{\bibfnamefont{N.}~\bibnamefont{Aghanim}} \bibnamefont{et~al.}
  (\bibinfo{collaboration}{Planck}), \bibinfo{journal}{Astron. Astrophys.}
  \textbf{\bibinfo{volume}{641}}, \bibinfo{pages}{A1}
  (\bibinfo{year}{2020}{\natexlab{a}}), \eprint{1807.06205}.

\bibitem[{\citenamefont{Aghanim et~al.}(2020{\natexlab{b}})}]{Planck:2018vyg}
\bibinfo{author}{\bibfnamefont{N.}~\bibnamefont{Aghanim}} \bibnamefont{et~al.}
  (\bibinfo{collaboration}{Planck}), \bibinfo{journal}{Astron. Astrophys.}
  \textbf{\bibinfo{volume}{641}}, \bibinfo{pages}{A6}
  (\bibinfo{year}{2020}{\natexlab{b}}), \bibinfo{note}{[Erratum:
  Astron.Astrophys. 652, C4 (2021)]}, \eprint{1807.06209}.

\bibitem[{\citenamefont{Aghanim et~al.}(2020{\natexlab{c}})}]{Planck:2019nip}
\bibinfo{author}{\bibfnamefont{N.}~\bibnamefont{Aghanim}} \bibnamefont{et~al.}
  (\bibinfo{collaboration}{Planck}), \bibinfo{journal}{Astron. Astrophys.}
  \textbf{\bibinfo{volume}{641}}, \bibinfo{pages}{A5}
  (\bibinfo{year}{2020}{\natexlab{c}}), \eprint{1907.12875}.

\bibitem[{\citenamefont{Tegmark et~al.}(2004)}]{SDSS:2003eyi}
\bibinfo{author}{\bibfnamefont{M.}~\bibnamefont{Tegmark}} \bibnamefont{et~al.}
  (\bibinfo{collaboration}{SDSS}), \bibinfo{journal}{Phys. Rev. D}
  \textbf{\bibinfo{volume}{69}}, \bibinfo{pages}{103501}
  (\bibinfo{year}{2004}), \eprint{astro-ph/0310723}.

\bibitem[{\citenamefont{Seljak et~al.}(2005)}]{SDSS:2004kqt}
\bibinfo{author}{\bibfnamefont{U.}~\bibnamefont{Seljak}} \bibnamefont{et~al.}
  (\bibinfo{collaboration}{SDSS}), \bibinfo{journal}{Phys. Rev. D}
  \textbf{\bibinfo{volume}{71}}, \bibinfo{pages}{103515}
  (\bibinfo{year}{2005}), \eprint{astro-ph/0407372}.

\bibitem[{\citenamefont{Perlmutter
  et~al.}(1999)}]{SupernovaCosmologyProject:1998vns}
\bibinfo{author}{\bibfnamefont{S.}~\bibnamefont{Perlmutter}}
  \bibnamefont{et~al.} (\bibinfo{collaboration}{Supernova Cosmology Project}),
  \bibinfo{journal}{Astrophys. J.} \textbf{\bibinfo{volume}{517}},
  \bibinfo{pages}{565} (\bibinfo{year}{1999}), \eprint{astro-ph/9812133}.

\bibitem[{\citenamefont{Riess et~al.}(1998)}]{SupernovaSearchTeam:1998fmf}
\bibinfo{author}{\bibfnamefont{A.~G.} \bibnamefont{Riess}} \bibnamefont{et~al.}
  (\bibinfo{collaboration}{Supernova Search Team}), \bibinfo{journal}{Astron.
  J.} \textbf{\bibinfo{volume}{116}}, \bibinfo{pages}{1009}
  (\bibinfo{year}{1998}), \eprint{astro-ph/9805201}.

\bibitem[{\citenamefont{Eisenstein et~al.}(2005)}]{SDSS:2005xqv}
\bibinfo{author}{\bibfnamefont{D.~J.} \bibnamefont{Eisenstein}}
  \bibnamefont{et~al.} (\bibinfo{collaboration}{SDSS}),
  \bibinfo{journal}{Astrophys. J.} \textbf{\bibinfo{volume}{633}},
  \bibinfo{pages}{560} (\bibinfo{year}{2005}), \eprint{astro-ph/0501171}.

\bibitem[{\citenamefont{Jain and Taylor}(2003)}]{Jain:2003tba}
\bibinfo{author}{\bibfnamefont{B.}~\bibnamefont{Jain}} \bibnamefont{and}
  \bibinfo{author}{\bibfnamefont{A.}~\bibnamefont{Taylor}},
  \bibinfo{journal}{Phys. Rev. Lett.} \textbf{\bibinfo{volume}{91}},
  \bibinfo{pages}{141302} (\bibinfo{year}{2003}), \eprint{astro-ph/0306046}.

\bibitem[{\citenamefont{Abbott et~al.}(2017)}]{LIGOScientific:2017adf}
\bibinfo{author}{\bibfnamefont{B.~P.} \bibnamefont{Abbott}}
  \bibnamefont{et~al.} (\bibinfo{collaboration}{LIGO Scientific, Virgo, 1M2H,
  Dark Energy Camera GW-E, DES, DLT40, Las Cumbres Observatory, VINROUGE,
  MASTER}), \bibinfo{journal}{Nature} \textbf{\bibinfo{volume}{551}},
  \bibinfo{pages}{85} (\bibinfo{year}{2017}), \eprint{1710.05835}.

\bibitem[{\citenamefont{Guth}(1981)}]{Guth:1980zm}
\bibinfo{author}{\bibfnamefont{A.~H.} \bibnamefont{Guth}},
  \bibinfo{journal}{Phys. Rev. D} \textbf{\bibinfo{volume}{23}},
  \bibinfo{pages}{347} (\bibinfo{year}{1981}).

\bibitem[{\citenamefont{Linde}(1982)}]{Linde:1981mu}
\bibinfo{author}{\bibfnamefont{A.~D.} \bibnamefont{Linde}},
  \bibinfo{journal}{Phys. Lett. B} \textbf{\bibinfo{volume}{108}},
  \bibinfo{pages}{389} (\bibinfo{year}{1982}).

\bibitem[{\citenamefont{Starobinsky}(1980)}]{Starobinsky:1980te}
\bibinfo{author}{\bibfnamefont{A.~A.} \bibnamefont{Starobinsky}},
  \bibinfo{journal}{Phys. Lett. B} \textbf{\bibinfo{volume}{91}},
  \bibinfo{pages}{99} (\bibinfo{year}{1980}).

\bibitem[{\citenamefont{Cai et~al.}(2010)\citenamefont{Cai, Saridakis, Setare,
  and Xia}}]{Cai:2009zp}
\bibinfo{author}{\bibfnamefont{Y.-F.} \bibnamefont{Cai}},
  \bibinfo{author}{\bibfnamefont{E.~N.} \bibnamefont{Saridakis}},
  \bibinfo{author}{\bibfnamefont{M.~R.} \bibnamefont{Setare}},
  \bibnamefont{and} \bibinfo{author}{\bibfnamefont{J.-Q.} \bibnamefont{Xia}},
  \bibinfo{journal}{Phys. Rept.} \textbf{\bibinfo{volume}{493}},
  \bibinfo{pages}{1} (\bibinfo{year}{2010}), \eprint{0909.2776}.

\bibitem[{\citenamefont{Battefeld and Peter}(2015)}]{Battefeld:2014uga}
\bibinfo{author}{\bibfnamefont{D.}~\bibnamefont{Battefeld}} \bibnamefont{and}
  \bibinfo{author}{\bibfnamefont{P.}~\bibnamefont{Peter}},
  \bibinfo{journal}{Phys. Rept.} \textbf{\bibinfo{volume}{571}},
  \bibinfo{pages}{1} (\bibinfo{year}{2015}), \eprint{1406.2790}.

\bibitem[{\citenamefont{Lyth and Wands}(2002)}]{Lyth:2001nq}
\bibinfo{author}{\bibfnamefont{D.~H.} \bibnamefont{Lyth}} \bibnamefont{and}
  \bibinfo{author}{\bibfnamefont{D.}~\bibnamefont{Wands}},
  \bibinfo{journal}{Phys. Lett. B} \textbf{\bibinfo{volume}{524}},
  \bibinfo{pages}{5} (\bibinfo{year}{2002}), \eprint{hep-ph/0110002}.

\bibitem[{\citenamefont{Brandenberger and Peter}(2017)}]{Brandenberger:2016vhg}
\bibinfo{author}{\bibfnamefont{R.}~\bibnamefont{Brandenberger}}
  \bibnamefont{and} \bibinfo{author}{\bibfnamefont{P.}~\bibnamefont{Peter}},
  \bibinfo{journal}{Found. Phys.} \textbf{\bibinfo{volume}{47}},
  \bibinfo{pages}{797} (\bibinfo{year}{2017}), \eprint{1603.05834}.

\bibitem[{\citenamefont{Bonanno}(2023)}]{Bonanno2023}
\bibinfo{author}{\bibfnamefont{A.}~\bibnamefont{Bonanno}},
  \emph{\bibinfo{title}{Asymptotic Safety and Cosmology}}
  (\bibinfo{publisher}{Springer Nature Singapore},
  \bibinfo{address}{Singapore}, \bibinfo{year}{2023}), pp.
  \bibinfo{pages}{1--27}, ISBN \bibinfo{isbn}{978-981-19-3079-9},
  \urlprefix\url{https://doi.org/10.1007/978-981-19-3079-9_23-1}.

\bibitem[{\citenamefont{Markov and Mukhanov}(1985)}]{Markov:1985py}
\bibinfo{author}{\bibfnamefont{M.~A.} \bibnamefont{Markov}} \bibnamefont{and}
  \bibinfo{author}{\bibfnamefont{V.~F.} \bibnamefont{Mukhanov}},
  \bibinfo{journal}{Nuovo Cim. B} \textbf{\bibinfo{volume}{86}},
  \bibinfo{pages}{97} (\bibinfo{year}{1985}).

\bibitem[{\citenamefont{Harko and Lobo}(2010)}]{Harko:2010mv}
\bibinfo{author}{\bibfnamefont{T.}~\bibnamefont{Harko}} \bibnamefont{and}
  \bibinfo{author}{\bibfnamefont{F.~S.~N.} \bibnamefont{Lobo}},
  \bibinfo{journal}{Eur. Phys. J. C} \textbf{\bibinfo{volume}{70}},
  \bibinfo{pages}{373} (\bibinfo{year}{2010}), \eprint{1008.4193}.

\bibitem[{\citenamefont{Lobo and Harko}(2015)}]{Lobo:2012af}
\bibinfo{author}{\bibfnamefont{F.~S.~N.} \bibnamefont{Lobo}} \bibnamefont{and}
  \bibinfo{author}{\bibfnamefont{T.}~\bibnamefont{Harko}}, in
  \emph{\bibinfo{booktitle}{{13th Marcel Grossmann Meeting on Recent
  Developments in Theoretical and Experimental General Relativity,
  Astrophysics, and Relativistic Field Theories}}} (\bibinfo{year}{2015}), pp.
  \bibinfo{pages}{1164--1166}, \eprint{1211.0426}.

\bibitem[{\citenamefont{Harko et~al.}(2011)\citenamefont{Harko, Lobo, Nojiri,
  and Odintsov}}]{Harko:2011kv}
\bibinfo{author}{\bibfnamefont{T.}~\bibnamefont{Harko}},
  \bibinfo{author}{\bibfnamefont{F.~S.~N.} \bibnamefont{Lobo}},
  \bibinfo{author}{\bibfnamefont{S.}~\bibnamefont{Nojiri}}, \bibnamefont{and}
  \bibinfo{author}{\bibfnamefont{S.~D.} \bibnamefont{Odintsov}},
  \bibinfo{journal}{Phys. Rev. D} \textbf{\bibinfo{volume}{84}},
  \bibinfo{pages}{024020} (\bibinfo{year}{2011}), \eprint{1104.2669}.

\bibitem[{\citenamefont{Moraes and Sahoo}(2017)}]{Moraes:2017zgm}
\bibinfo{author}{\bibfnamefont{P.~H. R.~S.} \bibnamefont{Moraes}}
  \bibnamefont{and} \bibinfo{author}{\bibfnamefont{P.~K.} \bibnamefont{Sahoo}},
  \bibinfo{journal}{Eur. Phys. J. C} \textbf{\bibinfo{volume}{77}},
  \bibinfo{pages}{480} (\bibinfo{year}{2017}), \eprint{1707.01360}.

\bibitem[{\citenamefont{Bertolami et~al.}(2008)\citenamefont{Bertolami, Lobo,
  and Paramos}}]{Bertolami:2008ab}
\bibinfo{author}{\bibfnamefont{O.}~\bibnamefont{Bertolami}},
  \bibinfo{author}{\bibfnamefont{F.~S.~N.} \bibnamefont{Lobo}},
  \bibnamefont{and} \bibinfo{author}{\bibfnamefont{J.}~\bibnamefont{Paramos}},
  \bibinfo{journal}{Phys. Rev. D} \textbf{\bibinfo{volume}{78}},
  \bibinfo{pages}{064036} (\bibinfo{year}{2008}), \eprint{0806.4434}.

\bibitem[{\citenamefont{Harko et~al.}(2018)\citenamefont{Harko, Koivisto, Lobo,
  Olmo, and Rubiera-Garcia}}]{Harko:2018gxr}
\bibinfo{author}{\bibfnamefont{T.}~\bibnamefont{Harko}},
  \bibinfo{author}{\bibfnamefont{T.~S.} \bibnamefont{Koivisto}},
  \bibinfo{author}{\bibfnamefont{F.~S.~N.} \bibnamefont{Lobo}},
  \bibinfo{author}{\bibfnamefont{G.~J.} \bibnamefont{Olmo}}, \bibnamefont{and}
  \bibinfo{author}{\bibfnamefont{D.}~\bibnamefont{Rubiera-Garcia}},
  \bibinfo{journal}{Phys. Rev. D} \textbf{\bibinfo{volume}{98}},
  \bibinfo{pages}{084043} (\bibinfo{year}{2018}), \eprint{1806.10437}.

\bibitem[{\citenamefont{Xu et~al.}(2019)\citenamefont{Xu, Li, Harko, and
  Liang}}]{Xu:2019sbp}
\bibinfo{author}{\bibfnamefont{Y.}~\bibnamefont{Xu}},
  \bibinfo{author}{\bibfnamefont{G.}~\bibnamefont{Li}},
  \bibinfo{author}{\bibfnamefont{T.}~\bibnamefont{Harko}}, \bibnamefont{and}
  \bibinfo{author}{\bibfnamefont{S.-D.} \bibnamefont{Liang}},
  \bibinfo{journal}{Eur. Phys. J. C} \textbf{\bibinfo{volume}{79}},
  \bibinfo{pages}{708} (\bibinfo{year}{2019}), \eprint{1908.04760}.

\bibitem[{\citenamefont{Bonanno and
  Reuter}(2002{\natexlab{a}})}]{Bonanno:2001xi}
\bibinfo{author}{\bibfnamefont{A.}~\bibnamefont{Bonanno}} \bibnamefont{and}
  \bibinfo{author}{\bibfnamefont{M.}~\bibnamefont{Reuter}},
  \bibinfo{journal}{Phys. Rev. D} \textbf{\bibinfo{volume}{65}},
  \bibinfo{pages}{043508} (\bibinfo{year}{2002}{\natexlab{a}}),
  \eprint{hep-th/0106133}.

\bibitem[{\citenamefont{Bonanno and
  Reuter}(2002{\natexlab{b}})}]{Bonanno:2001hi}
\bibinfo{author}{\bibfnamefont{A.}~\bibnamefont{Bonanno}} \bibnamefont{and}
  \bibinfo{author}{\bibfnamefont{M.}~\bibnamefont{Reuter}},
  \bibinfo{journal}{Phys. Lett. B} \textbf{\bibinfo{volume}{527}},
  \bibinfo{pages}{9} (\bibinfo{year}{2002}{\natexlab{b}}),
  \eprint{astro-ph/0106468}.

\bibitem[{\citenamefont{Cai and Easson}(2011)}]{Cai:2011kd}
\bibinfo{author}{\bibfnamefont{Y.-F.} \bibnamefont{Cai}} \bibnamefont{and}
  \bibinfo{author}{\bibfnamefont{D.~A.} \bibnamefont{Easson}},
  \bibinfo{journal}{Phys. Rev. D} \textbf{\bibinfo{volume}{84}},
  \bibinfo{pages}{103502} (\bibinfo{year}{2011}), \eprint{1107.5815}.

\bibitem[{\citenamefont{Bonanno}(2012)}]{Bonanno:2012jy}
\bibinfo{author}{\bibfnamefont{A.}~\bibnamefont{Bonanno}},
  \bibinfo{journal}{Phys. Rev. D} \textbf{\bibinfo{volume}{85}},
  \bibinfo{pages}{081503} (\bibinfo{year}{2012}), \eprint{1203.1962}.

\bibitem[{\citenamefont{Bonanno and Saueressig}(2017)}]{Bonanno:2017pkg}
\bibinfo{author}{\bibfnamefont{A.}~\bibnamefont{Bonanno}} \bibnamefont{and}
  \bibinfo{author}{\bibfnamefont{F.}~\bibnamefont{Saueressig}},
  \bibinfo{journal}{Comptes Rendus Physique} \textbf{\bibinfo{volume}{18}},
  \bibinfo{pages}{254} (\bibinfo{year}{2017}), \eprint{1702.04137}.

\bibitem[{\citenamefont{Bonanno et~al.}(2018)\citenamefont{Bonanno, Platania,
  and Saueressig}}]{Bonanno:2018gck}
\bibinfo{author}{\bibfnamefont{A.}~\bibnamefont{Bonanno}},
  \bibinfo{author}{\bibfnamefont{A.}~\bibnamefont{Platania}}, \bibnamefont{and}
  \bibinfo{author}{\bibfnamefont{F.}~\bibnamefont{Saueressig}},
  \bibinfo{journal}{Phys. Lett. B} \textbf{\bibinfo{volume}{784}},
  \bibinfo{pages}{229} (\bibinfo{year}{2018}), \eprint{1803.02355}.

\bibitem[{\citenamefont{Pawlowski and Reichert}(2021)}]{Pawlowski:2020qer}
\bibinfo{author}{\bibfnamefont{J.~M.} \bibnamefont{Pawlowski}}
  \bibnamefont{and} \bibinfo{author}{\bibfnamefont{M.}~\bibnamefont{Reichert}},
  \bibinfo{journal}{Front. in Phys.} \textbf{\bibinfo{volume}{8}},
  \bibinfo{pages}{551848} (\bibinfo{year}{2021}), \eprint{2007.10353}.

\bibitem[{\citenamefont{Bonanno et~al.}(2022)\citenamefont{Bonanno, Denz,
  Pawlowski, and Reichert}}]{Bonanno:2021squ}
\bibinfo{author}{\bibfnamefont{A.}~\bibnamefont{Bonanno}},
  \bibinfo{author}{\bibfnamefont{T.}~\bibnamefont{Denz}},
  \bibinfo{author}{\bibfnamefont{J.~M.} \bibnamefont{Pawlowski}},
  \bibnamefont{and} \bibinfo{author}{\bibfnamefont{M.}~\bibnamefont{Reichert}},
  \bibinfo{journal}{SciPost Phys.} \textbf{\bibinfo{volume}{12}},
  \bibinfo{pages}{001} (\bibinfo{year}{2022}), \eprint{2102.02217}.

\bibitem[{\citenamefont{Bonanno et~al.}(2023)\citenamefont{Bonanno, Malafarina,
  and Panassiti}}]{Bonanno:2023rzk}
\bibinfo{author}{\bibfnamefont{A.}~\bibnamefont{Bonanno}},
  \bibinfo{author}{\bibfnamefont{D.}~\bibnamefont{Malafarina}},
  \bibnamefont{and} \bibinfo{author}{\bibfnamefont{A.}~\bibnamefont{Panassiti}}
  (\bibinfo{year}{2023}), \eprint{2308.10890}.

\bibitem[{\citenamefont{Reuter}(1998)}]{Reuter:1996cp}
\bibinfo{author}{\bibfnamefont{M.}~\bibnamefont{Reuter}},
  \bibinfo{journal}{Phys. Rev. D} \textbf{\bibinfo{volume}{57}},
  \bibinfo{pages}{971} (\bibinfo{year}{1998}), \eprint{hep-th/9605030}.

\bibitem[{\citenamefont{Weinberg}(1980)}]{Weinberg:1980gg}
\bibinfo{author}{\bibfnamefont{S.}~\bibnamefont{Weinberg}},
  \emph{\bibinfo{title}{{ULTRAVIOLET DIVERGENCES IN QUANTUM THEORIES OF
  GRAVITATION}}} (\bibinfo{publisher}{{Cambridge University Press}},
  \bibinfo{year}{1980}), pp. \bibinfo{pages}{790--831}.

\bibitem[{\citenamefont{Eichhorn}(2019)}]{Eichhorn:2018yfc}
\bibinfo{author}{\bibfnamefont{A.}~\bibnamefont{Eichhorn}},
  \bibinfo{journal}{Front. Astron. Space Sci.} \textbf{\bibinfo{volume}{5}},
  \bibinfo{pages}{47} (\bibinfo{year}{2019}), \eprint{1810.07615}.

\bibitem[{\citenamefont{Reichert}(2020)}]{Reichert:2020mja}
\bibinfo{author}{\bibfnamefont{M.}~\bibnamefont{Reichert}},
  \bibinfo{journal}{PoS} \textbf{\bibinfo{volume}{384}}, \bibinfo{pages}{005}
  (\bibinfo{year}{2020}).

\bibitem[{\citenamefont{Reuter and Saueressig}(2019)}]{Reuter:2019byg}
\bibinfo{author}{\bibfnamefont{M.}~\bibnamefont{Reuter}} \bibnamefont{and}
  \bibinfo{author}{\bibfnamefont{F.}~\bibnamefont{Saueressig}},
  \emph{\bibinfo{title}{{Quantum Gravity and the Functional Renormalization
  Group}: {The Road towards Asymptotic Safety}}} (\bibinfo{publisher}{Cambridge
  University Press}, \bibinfo{year}{2019}), ISBN
  \bibinfo{isbn}{978-1-107-10732-8, 978-1-108-67074-6}.

\bibitem[{\citenamefont{Percacci}(2017)}]{Percacci:2017fkn}
\bibinfo{author}{\bibfnamefont{R.}~\bibnamefont{Percacci}},
  \emph{\bibinfo{title}{{An Introduction to Covariant Quantum Gravity and
  Asymptotic Safety}}}, vol.~\bibinfo{volume}{3} of \emph{\bibinfo{series}{100
  Years of General Relativity}} (\bibinfo{publisher}{World Scientific},
  \bibinfo{year}{2017}), ISBN \bibinfo{isbn}{978-981-320-717-2,
  978-981-320-719-6}.

\bibitem[{\citenamefont{Bonanno
  et~al.}(2020{\natexlab{a}})\citenamefont{Bonanno, Casadio, and
  Platania}}]{Bonanno:2019ilz}
\bibinfo{author}{\bibfnamefont{A.}~\bibnamefont{Bonanno}},
  \bibinfo{author}{\bibfnamefont{R.}~\bibnamefont{Casadio}}, \bibnamefont{and}
  \bibinfo{author}{\bibfnamefont{A.}~\bibnamefont{Platania}},
  \bibinfo{journal}{JCAP} \textbf{\bibinfo{volume}{01}}, \bibinfo{pages}{022}
  (\bibinfo{year}{2020}{\natexlab{a}}), \eprint{1910.11393}.

\bibitem[{\citenamefont{Bonanno
  et~al.}(2020{\natexlab{b}})\citenamefont{Bonanno, Eichhorn, Gies, Pawlowski,
  Percacci, Reuter, Saueressig, and Vacca}}]{Bonanno:2020bil}
\bibinfo{author}{\bibfnamefont{A.}~\bibnamefont{Bonanno}},
  \bibinfo{author}{\bibfnamefont{A.}~\bibnamefont{Eichhorn}},
  \bibinfo{author}{\bibfnamefont{H.}~\bibnamefont{Gies}},
  \bibinfo{author}{\bibfnamefont{J.~M.} \bibnamefont{Pawlowski}},
  \bibinfo{author}{\bibfnamefont{R.}~\bibnamefont{Percacci}},
  \bibinfo{author}{\bibfnamefont{M.}~\bibnamefont{Reuter}},
  \bibinfo{author}{\bibfnamefont{F.}~\bibnamefont{Saueressig}},
  \bibnamefont{and} \bibinfo{author}{\bibfnamefont{G.~P.} \bibnamefont{Vacca}},
  \bibinfo{journal}{Front. in Phys.} \textbf{\bibinfo{volume}{8}},
  \bibinfo{pages}{269} (\bibinfo{year}{2020}{\natexlab{b}}),
  \eprint{2004.06810}.

\bibitem[{\citenamefont{Platania}(2019)}]{Platania:2019kyx}
\bibinfo{author}{\bibfnamefont{A.}~\bibnamefont{Platania}},
  \bibinfo{journal}{Eur. Phys. J. C} \textbf{\bibinfo{volume}{79}},
  \bibinfo{pages}{470} (\bibinfo{year}{2019}), \eprint{1903.10411}.

\bibitem[{\citenamefont{Zel'dovich}(1961)}]{Zeldovich:1961sbr}
\bibinfo{author}{\bibfnamefont{Y.~B.} \bibnamefont{Zel'dovich}},
  \bibinfo{journal}{Zh. Eksp. Teor. Fiz.} \textbf{\bibinfo{volume}{41}},
  \bibinfo{pages}{1609} (\bibinfo{year}{1961}).

\bibitem[{\citenamefont{Zeldovich}(1972)}]{Zeldovich:1972zz}
\bibinfo{author}{\bibfnamefont{Y.~B.} \bibnamefont{Zeldovich}},
  \bibinfo{journal}{Mon. Not. Roy. Astron. Soc.}
  \textbf{\bibinfo{volume}{160}}, \bibinfo{pages}{1P} (\bibinfo{year}{1972}).

\bibitem[{\citenamefont{Chavanis}(2015)}]{Chavanis:2014lra}
\bibinfo{author}{\bibfnamefont{P.-H.} \bibnamefont{Chavanis}},
  \bibinfo{journal}{Phys. Rev. D} \textbf{\bibinfo{volume}{92}},
  \bibinfo{pages}{103004} (\bibinfo{year}{2015}), \eprint{1412.0743}.

\bibitem[{\citenamefont{Baumann}(2022)}]{Baumann:2022mni}
\bibinfo{author}{\bibfnamefont{D.}~\bibnamefont{Baumann}},
  \emph{\bibinfo{title}{{Cosmology}}} (\bibinfo{publisher}{Cambridge University
  Press}, \bibinfo{year}{2022}), ISBN \bibinfo{isbn}{978-1-108-93709-2,
  978-1-108-83807-8}.

\bibitem[{\citenamefont{Lyth and Riotto}(1999)}]{Lyth:1998xn}
\bibinfo{author}{\bibfnamefont{D.~H.} \bibnamefont{Lyth}} \bibnamefont{and}
  \bibinfo{author}{\bibfnamefont{A.}~\bibnamefont{Riotto}},
  \bibinfo{journal}{Phys. Rept.} \textbf{\bibinfo{volume}{314}},
  \bibinfo{pages}{1} (\bibinfo{year}{1999}), \eprint{hep-ph/9807278}.

\bibitem[{\citenamefont{Chen et~al.}(2013)\citenamefont{Chen, Firouzjahi,
  Namjoo, and Sasaki}}]{Chen:2013kta}
\bibinfo{author}{\bibfnamefont{X.}~\bibnamefont{Chen}},
  \bibinfo{author}{\bibfnamefont{H.}~\bibnamefont{Firouzjahi}},
  \bibinfo{author}{\bibfnamefont{M.~H.} \bibnamefont{Namjoo}},
  \bibnamefont{and} \bibinfo{author}{\bibfnamefont{M.}~\bibnamefont{Sasaki}},
  \bibinfo{journal}{JCAP} \textbf{\bibinfo{volume}{09}}, \bibinfo{pages}{012}
  (\bibinfo{year}{2013}), \eprint{1306.2901}.

\bibitem[{\citenamefont{Ray}(1972)}]{Ray:1972}
\bibinfo{author}{\bibfnamefont{J.~R.} \bibnamefont{Ray}}, \bibinfo{journal}{J.
  Math. Phys.} \textbf{\bibinfo{volume}{13}}, \bibinfo{pages}{1451}
  (\bibinfo{year}{1972}).

\bibitem[{\citenamefont{Gordon and Hu}(2004)}]{Gordon:2004ez}
\bibinfo{author}{\bibfnamefont{C.}~\bibnamefont{Gordon}} \bibnamefont{and}
  \bibinfo{author}{\bibfnamefont{W.}~\bibnamefont{Hu}}, \bibinfo{journal}{Phys.
  Rev. D} \textbf{\bibinfo{volume}{70}}, \bibinfo{pages}{083003}
  (\bibinfo{year}{2004}), \eprint{astro-ph/0406496}.

\bibitem[{\citenamefont{Sasaki}(1986)}]{Sasaki:1986hm}
\bibinfo{author}{\bibfnamefont{M.}~\bibnamefont{Sasaki}},
  \bibinfo{journal}{Prog. Theor. Phys.} \textbf{\bibinfo{volume}{76}},
  \bibinfo{pages}{1036} (\bibinfo{year}{1986}).

\bibitem[{\citenamefont{Mukhanov}(1988)}]{Mukhanov:1988jd}
\bibinfo{author}{\bibfnamefont{V.~F.} \bibnamefont{Mukhanov}},
  \bibinfo{journal}{Sov. Phys. JETP} \textbf{\bibinfo{volume}{67}},
  \bibinfo{pages}{1297} (\bibinfo{year}{1988}).

\bibitem[{\citenamefont{Bunch and Davies}(1978)}]{Bunch:1978yq}
\bibinfo{author}{\bibfnamefont{T.~S.} \bibnamefont{Bunch}} \bibnamefont{and}
  \bibinfo{author}{\bibfnamefont{P.~C.~W.} \bibnamefont{Davies}},
  \bibinfo{journal}{Proc. Roy. Soc. Lond. A} \textbf{\bibinfo{volume}{360}},
  \bibinfo{pages}{117} (\bibinfo{year}{1978}).

\bibitem[{\citenamefont{Ade et~al.}(2018)}]{BICEP2:2018kqh}
\bibinfo{author}{\bibfnamefont{P.~A.~R.} \bibnamefont{Ade}}
  \bibnamefont{et~al.} (\bibinfo{collaboration}{BICEP2, Keck Array}),
  \bibinfo{journal}{Phys. Rev. Lett.} \textbf{\bibinfo{volume}{121}},
  \bibinfo{pages}{221301} (\bibinfo{year}{2018}), \eprint{1810.05216}.

\bibitem[{\citenamefont{Aiola et~al.}(2020)}]{ACT:2020gnv}
\bibinfo{author}{\bibfnamefont{S.}~\bibnamefont{Aiola}} \bibnamefont{et~al.}
  (\bibinfo{collaboration}{ACT}), \bibinfo{journal}{JCAP}
  \textbf{\bibinfo{volume}{12}}, \bibinfo{pages}{047} (\bibinfo{year}{2020}),
  \eprint{2007.07288}.

\bibitem[{\citenamefont{Forconi et~al.}(2021)\citenamefont{Forconi, Giar\`e,
  Di~Valentino, and Melchiorri}}]{Forconi:2021que}
\bibinfo{author}{\bibfnamefont{M.}~\bibnamefont{Forconi}},
  \bibinfo{author}{\bibfnamefont{W.}~\bibnamefont{Giar\`e}},
  \bibinfo{author}{\bibfnamefont{E.}~\bibnamefont{Di~Valentino}},
  \bibnamefont{and}
  \bibinfo{author}{\bibfnamefont{A.}~\bibnamefont{Melchiorri}},
  \bibinfo{journal}{Phys. Rev. D} \textbf{\bibinfo{volume}{104}},
  \bibinfo{pages}{103528} (\bibinfo{year}{2021}), \eprint{2110.01695}.

\bibitem[{\citenamefont{Dutcher et~al.}(2021)}]{SPT-3G:2021eoc}
\bibinfo{author}{\bibfnamefont{D.}~\bibnamefont{Dutcher}} \bibnamefont{et~al.}
  (\bibinfo{collaboration}{SPT-3G}), \bibinfo{journal}{Phys. Rev. D}
  \textbf{\bibinfo{volume}{104}}, \bibinfo{pages}{022003}
  (\bibinfo{year}{2021}), \eprint{2101.01684}.

\bibitem[{\citenamefont{de~Cesare et~al.}(2016)\citenamefont{de~Cesare, Lizzi,
  and Sakellariadou}}]{deCesare:2016dnp}
\bibinfo{author}{\bibfnamefont{M.}~\bibnamefont{de~Cesare}},
  \bibinfo{author}{\bibfnamefont{F.}~\bibnamefont{Lizzi}}, \bibnamefont{and}
  \bibinfo{author}{\bibfnamefont{M.}~\bibnamefont{Sakellariadou}},
  \bibinfo{journal}{Phys. Lett. B} \textbf{\bibinfo{volume}{760}},
  \bibinfo{pages}{498} (\bibinfo{year}{2016}), \eprint{1603.04170}.

\bibitem[{\citenamefont{Abazajian et~al.}(2022)}]{CMB-S4:2020lpa}
\bibinfo{author}{\bibfnamefont{K.}~\bibnamefont{Abazajian}}
  \bibnamefont{et~al.} (\bibinfo{collaboration}{CMB-S4}),
  \bibinfo{journal}{Astrophys. J.} \textbf{\bibinfo{volume}{926}},
  \bibinfo{pages}{54} (\bibinfo{year}{2022}), \eprint{2008.12619}.

\bibitem[{\citenamefont{Zegeye et~al.}(2023)}]{CMB-S4:2023zem}
\bibinfo{author}{\bibfnamefont{D.}~\bibnamefont{Zegeye}} \bibnamefont{et~al.}
  (\bibinfo{collaboration}{CMB-S4}), \bibinfo{journal}{Phys. Rev. D}
  \textbf{\bibinfo{volume}{108}}, \bibinfo{pages}{103536}
  (\bibinfo{year}{2023}), \eprint{2303.00916}.

\bibitem[{\citenamefont{Allys et~al.}(2023)}]{LiteBIRD:2022cnt}
\bibinfo{author}{\bibfnamefont{E.}~\bibnamefont{Allys}} \bibnamefont{et~al.}
  (\bibinfo{collaboration}{LiteBIRD}), \bibinfo{journal}{PTEP}
  \textbf{\bibinfo{volume}{2023}}, \bibinfo{pages}{042F01}
  (\bibinfo{year}{2023}), \eprint{2202.02773}.

\bibitem[{\citenamefont{Delabrouille et~al.}(2018)}]{CORE:2017oje}
\bibinfo{author}{\bibfnamefont{J.}~\bibnamefont{Delabrouille}}
  \bibnamefont{et~al.} (\bibinfo{collaboration}{CORE}), \bibinfo{journal}{JCAP}
  \textbf{\bibinfo{volume}{04}}, \bibinfo{pages}{014} (\bibinfo{year}{2018}),
  \eprint{1706.04516}.

\bibitem[{\citenamefont{Adak et~al.}(2022)\citenamefont{Adak, Sen, Basak,
  Delabrouille, Ghosh, Rotti, Mart\'\i{}nez-Solaeche, and
  Souradeep}}]{Adak:2021lbu}
\bibinfo{author}{\bibfnamefont{D.}~\bibnamefont{Adak}},
  \bibinfo{author}{\bibfnamefont{A.}~\bibnamefont{Sen}},
  \bibinfo{author}{\bibfnamefont{S.}~\bibnamefont{Basak}},
  \bibinfo{author}{\bibfnamefont{J.}~\bibnamefont{Delabrouille}},
  \bibinfo{author}{\bibfnamefont{T.}~\bibnamefont{Ghosh}},
  \bibinfo{author}{\bibfnamefont{A.}~\bibnamefont{Rotti}},
  \bibinfo{author}{\bibfnamefont{G.}~\bibnamefont{Mart\'\i{}nez-Solaeche}},
  \bibnamefont{and}
  \bibinfo{author}{\bibfnamefont{T.}~\bibnamefont{Souradeep}},
  \bibinfo{journal}{Mon. Not. Roy. Astron. Soc.}
  \textbf{\bibinfo{volume}{514}}, \bibinfo{pages}{3002} (\bibinfo{year}{2022}),
  \eprint{2110.12362}.

\end{thebibliography}

\end{document}